\documentclass[a4paper,aps,prd,twocolumn,showkeys,showpacs,nofootinbib]{revtex4}
\usepackage{latexsym}
\usepackage{amsbsy}
\usepackage{mathrsfs}
\usepackage{color}
\usepackage{psfrag}
\usepackage{enumerate}
\usepackage{amsmath,amssymb,calc,amsfonts}
\usepackage{graphicx,calc,epsfig}
\usepackage[multiple]{footmisc}
\usepackage[english]{babel} 
\usepackage{mathtools}
\usepackage{bbm}
\usepackage{accents}
\usepackage{dsfont}
\usepackage{rotating}
\usepackage{bbold}

\def\ut#1{\rlap{\lower1ex\hbox{$\sim$}}#1{}}

\newcommand{\be}{\nopagebreak[3]\begin{equation}}
\newcommand{\ee}{\end{equation}}
\newcommand{\ba}{\nopagebreak[3]\begin{eqnarray}}
\newcommand{\ea}{\end{eqnarray}}
\DeclareFontFamily{U}{rsfs}{}         
\DeclareFontShape{U}{rsfs}{m}{n}{<5> rsfs5 <6><7> rsfs7          
  <8><9><10><10.95><12><14.4><17.28><20.74><24.88> rsfs10}{}     
\DeclareMathAlphabet{\mathfs}{U}{rsfs}{m}{n}                     
                               
\newcommand{\tr}{\mathrm{tr}}

\def\pb#1{\rlap{\lower1.5ex\hbox{$\longleftarrow$}}{#1}}
\def\dpb#1{\rlap{\lower1.5ex\hbox{$\Longleftarrow$}}{#1}}
\def\spb#1{\rlap{\lower1.5ex\hbox{$\leftarrow$}}{#1}}
\def\sdpb#1{\rlap{\lower1.5ex\hbox{$\Leftarrow$}}{#1}}

\definecolor{blue}{rgb}{0,0,1}
\definecolor{green}{rgb}{0,1,0}
\definecolor{red}{rgb}{1,0,0}
\definecolor{vio}{rgb}{1,0,1}
\definecolor{ama}{rgb}{1,1,0}
\usepackage[all]{xy}
\xyoption{knot}
\xyoption{frame}
\xyoption{arc}
\xyoption{curve}
\xyoption{poly}
\usepackage{hyperref}
\hypersetup{
  colorlinks   = true, 
  urlcolor     = blue, 
  linkcolor    = blue, 
  citecolor   =  red 
}
\def\be{\begin{equation}}
\def\ee{\end{equation}}
\def\ba{\begin{eqnarray}}
\def\ea{\end{eqnarray}}

\begin{document}

\begin{flushleft}
KCL-PH-TH/2016-66
\end{flushleft}

\title{Relational evolution of effectively interacting GFT
quantum gravity condensates}

\date{\today}

\author{Andreas G. A. Pithis$^1$}\email{andreas.pithis@kcl.ac.uk}
\author{Mairi Sakellariadou$^{1,2}$}\email{mairi.sakellariadou@kcl.ac.uk}
\affiliation{$^1$Theoretical Particle Physics and Cosmology Group, Department of Physics, King's College London, University of London, Strand, London, WC2R 2LS, U.K.\\
$^2$Perimeter Institute for Theoretical Physics, Waterloo, Ontario N2L 2Y5, Canada}

\thanks{}

\begin{abstract}

We study the impact of effective interactions onto relationally evolving GFT condensates based on real-valued fields. In a first step we show that a free condensate configuration in an isotropic restriction settles dynamically into a low-spin configuration of the quantum geometry. This goes hand in hand with the accelerated and exponential expansion of its volume, as well as the vanishing of its relative uncertainty which suggests the classicalization of the quantum geometry. The dynamics of the emergent space can then be given in terms of the classical Friedmann equations. In contrast to models based on complex-valued fields, solutions avoiding the singularity problem can only be found if the initial conditions are appropriately chosen. We then turn to the analysis of the influence of effective interactions on the dynamics by studying in particular the Thomas-Fermi regime. In this context, at the cost of fine-tuning, an epoch of inflationary expansion of quantum geometric origin can be implemented. Finally, and for the first time, we study anisotropic GFT condensate configurations and show that such systems tend to isotropize quickly as the value of the relational clock grows. This paves the way to a more systematic investigation of anisotropies in the context of GFT condensate cosmology.

\end{abstract}


\maketitle

\section{Introduction}

The hardest challenge for discrete quantum gravity approaches which rely on pregeometric structures of \textit{a priori} nonspatio-temporal nature is to retrieve, within an appropriate limit, the geometry of continuum spacetimes and its dynamics given by General Relativity. A proposal to recover the classical general realtivistic picture was given e.g. in Ref. \cite{Geometrogenesis}, where it was suggested that a phase transition in an underlying miscroscopic quantum gravity system could connect a discrete pregeometric with a continuum geometric phase. 

In the Group Field Theory (GFT) approach to quantum gravity \cite{GFT}, itself a second quantization formulation of Loop Quantum Gravity (LQG) \cite{LQG,GFTLQG}, this idea is adopted and it is conjectured that the process responsible for the emergence of a continuum geometric phase is given by a condensation of bosonic GFT quanta, each representing an atom of space \cite{GFTGeometrogenesis}. This hypothesis finds rising support through Functional Renormalization Group \cite{FRG} analyses of increasingly sophisticated GFT models \cite{TGFTFRG,FRGcurvedspace,TMRG,GFTRGReview}. In all cases considered so far, indications are found for a phase transition separating a symmetric from a broken (condensate) phase as the``mass parameter'' changes its sign to negative values in the IR limit of the theory.

The basic idea of the GFT based quantum cosmology spin-off is to identify this condensate phase with a continuum spacetime and to derive, through mean field techniques for the corresponding GFT condensate states, an effective dynamics with a tentative cosmological interpretation. This strategy has led to promising results in a quickly growing body of work \cite{GFC,GFCExample,GFCExample2,GFCOthers,GFCEmergentFriedmann,GFCEmergentFriedmann2,
GFClowspin,GFClowspinStaticEffInt,GFCEmergentFriedmann3,GFCReview} of which the recovery of the Friedmann dynamics for the emergent homogeneous and isotropic geometry is perhaps the most striking one \cite{GFCEmergentFriedmann}. This lends strong support to the idea that GFT condensate states are appropriate for studying the cosmological sector of LQG.

Within this context, the purpose of this article is to extend previous work by the same authors on the subject of GFT condensate cosmology presented in Ref. \cite{GFClowspinStaticEffInt}. In the latter, basing the analysis on real-valued GFT fields, it was shown for free and effectively interacting static GFT condensate systems in an isotropic restriction that the configurations indeed consist of many smallest discrete and almost flat building blocks of the quantum geometry giving rise to an effectively continuous $3$-geometry. This suggests that the interpretation of the condensate phase as corresponding to a geometric phase is a sensible conjecture and encourages us to study the relational evolution of such condensate systems. To do so, we incorporate the concept of a relational clock introduced to the context of GFT condensate cosmology in Ref. \cite{GFCEmergentFriedmann}, allowing us to extract information about the dynamics of the emergent $3$-geometries. Toward the end of this article, we will also lift the isotropic restriction to study more general, i.e., anisotropic free and effectively interacting condensates. This will allow for a more systematic exploration of anisotropic condensate models and their relation to models within, for instance, Loop Quantum Cosmology (LQC) \cite{LQC}.

To this aim, the article is organized as follows: In Section \ref{genset}, we start with a short recapitulation of basics of the GFT approach to quantum gravity in subsection \ref{gftrecap} and then briefly motivate its quantum cosmology spin-off in subsection \ref{gfcrecap}. The reader who is familiar with these concepts is invited to proceed directly to Section \ref{sectionstatic}. There, we first of all review some of the techniques developed in Ref. \cite{GFClowspinStaticEffInt} in order to extract the quantum geometric information encoded by a static free and isotropic condensate field\footnote{The isotropic restriction considered here reduces the number of coordinates parametrizing the field domain which is to some extent different from the one employed in Ref. \cite{GFCEmergentFriedmann} where the number of spin labels is reduced to one $j$.} and then further elaborate on this point by computing its noncommutative Fourier transform \cite{NCFT}. In Section \ref{relationalevolutioniso} we study the relational evolution of such a system. In particular, in the free case in subsections \ref{lowspinisotropic} and \ref{emergentfriedmanndynamics} it is shown that the corresponding quantum geometry settles quickly into the lowest nontrivial configuration available to the system and that the emergent geometry obeys Friedmann-like dynamics which is akin to the results shown in Refs. \cite{GFCEmergentFriedmann,GFCEmergentFriedmann2,GFClowspin}. In addition, it is also demonstrated that the relative uncertainty of the expectation value of the volume vanishes at late times which suggests the classicalization of the quantum geometry. For a particular choice of initial conditions, solutions avoiding the singularity problem can be found. We investigate the influence of effective interactions onto the condensate system in subsections \ref{isotropicthomasfermi}, \ref{isotropicacceleratedexpansion} and \ref{isotropiclinearizedsystem} where we start with the Thomas-Fermi regime and then proceed to a formal stability analysis of the full dynamical system using techniques of stability theory. In the former it is possible to demonstrate in close analogy to Ref. \cite{GFCEmergentFriedmann3} that for a particular choice of interaction terms one can accommodate for an era of inflationary expansion. In Section \ref{anisotropies} we investigate how anisotropic condensate configurations can be constructed and then show in subsection \ref{anisoiso},  how they dynamically isotropize. Conversely, it is shown that anisotropic contributions to the condensate dominate towards small volumes. The explanation for this particular behavior is given by means of a formal stability analysis of the anisotropic system in \ref{anisostability}. In Section \ref{Discussion} we summarize our results, discuss limitations of our analysis and propose directions for future research.

In Appendix \ref{AppendixA} we supplement the second section of this article by further discussing the notion of the noncommutative Fourier transform for GFT fields and how the $3$-metric can in principle be reconstructed from it. 

\section{General setup}\label{genset}

\subsection{Group Field Theory}\label{gftrecap}

Group Field Theories are QFTs defined over group manifolds with combinatorially nonlocal interaction terms intending to generalize matrix models for $2$d quantum gravity to higher dimensions \cite{GFT,MM}. The basic idea of this approach is that all data encoded in the fields is exclusively of combinatorial and algebraic nature which, by construction, turns GFT into a manifestly background independent and generally  covariant field theoretic framework.

More specifically, the fields are defined over the Lie group $G$ (or dually on its associated Lie algebra $\mathfrak{g}$) which represents the local gauge group of gravity. For models which aim at providing a second quantized reformulation of the kinematics and dynamics of canonical LQG, $G=\mathrm{SU}(2)$ which is the gauge group of Ashtekar-Barbero gravity. In the following discussion we stick to this choice and refer the reader to Refs. \cite{GFT,GFTLQG,GFTReview, GFT4All} for more details.

By selecting a type of field and its action, one specifies the classical GFT. For $4$d quantum gravity models, one chooses a complex-valued scalar field living on $d=4$ copies of $G$, i.e.,
\be
\varphi(g_I):G^d\to\mathbb{C}
\ee
where $I=1,...,d$. The group elements $g_I$ correspond to parallel transports $\mathcal{P}e^{\int_{e_I}A}$ of the gravitational connection $A$ along a link $e_I$. A further ingredient is the imposition of invariance of the field under the right action of $G$, i.e.,
\be\label{Rightinvariance}
\varphi(g_1 h,...,g_d h)=\varphi(g_1,...,g_d),~~~\forall h\in G
\ee
implying gauge invariance at the vertex from which the $d$ links emanate. In the equivalent formulation of the theory, where the field lives on the noncommutative momentum space $\mathfrak{g}^d$, this right invariance translates into a closure of the $d$ faces dual to the linke $e_I$ to form a tetrahedron or $3$-simplex, as further motivated in Appendix \ref{AppendixA}.

The action is then most generally given by 
\be\label{GFTaction}
S[\varphi,\bar{\varphi}]=\int_G(dg)^d\int_G(dg')^d\bar{\varphi}(g_I)\mathcal{K}(g_I,g'_I)\varphi(g_I)+\mathcal{V}
\ee
with $dg$ denoting the normalized Haar measure on $G$. The kinetic kernel $\mathcal{K}$ is typically local whereas the interaction term $\mathcal{V}[\varphi,\bar{\varphi}]$ is in general a nonlinear and nonlocal convolution of the GFT field with itself. From the action one obtains the classical equation of motion of the field, given by
\be
\int_G(dg')^d\mathcal{K}(g'_I,g_I)\varphi(g_I)+\frac{\delta \mathcal{V}}{\delta\bar{\varphi}(g_I)}=0.
\ee

Moving over to the quantum theory, the dynamics is defined in terms of the partition function
\be
Z=\int[\mathcal{D}\varphi][\mathcal{D}\bar{\varphi}]e^{-S[\varphi,\bar{\varphi}]}.
\ee
In the perturbative expansion in terms of the coupling constant(s) defined in $\mathcal{V}$ over the Fock vacuum one sees that the GFT Feynman diagrams are dual to cellular complexes of arbitrary topology if the field arguments in $\mathcal{V}$ are related to one another by means of a specific combinatorially nonlocal pattern \cite{GFT}. In particular, a so-called simplicial interaction term pairs five copies of the field in such a way that their corresponding $3$-simplices are glued together to form a $4$-simplex. 

Typically, a partition function for $4$d quantum gravity is constructed from the path integral quantization of the so-called Ooguri model \cite{Ooguri}. In turn, this defines a quantization of BF-theory in $4$d which is a topological field theory. In this model, the corresponding GFT field lives on four copies of $G=\mathrm{Spin}(4)$ or $\mathrm{SL}(2,\mathbb{C})$ and the action is specified by an ultralocal kinetic kernel and the aforementioned simplicial interaction term. To construct a unique geometry for the simplicial complex, so-called \textit{simplicity constraints} have to be imposed which reduce the nongeometric topological theory to the gravitational sector. More specifically, when invoking the correspondence between GFT and spin foam models \cite{GFT,GFTSF}, it is possible to show that the GFT quantization of the Ooguri model with properly imposed simplicity constraints, which i.a.  reduce $G$ to $\mathrm{SU}(2)$ on the boundary \cite{SFSC,GFTSG}, corresponds to a spin foam model for quantum gravity giving in turn a covariant QFT formulation of the dynamics of LQG \cite{LQG,SF}.     

In LQG, boundary spin network states of the spin foam correspond to discrete quantum $3$-geometries \cite{LQG,LQGdiscreteness}. From the point of view that GFT provides a second quantization of the kinematics of LQG, these boundary states are elements of the GFT Fock space \cite{GFTLQG}. In other words, in this picture open spin network vertices or their dual quantum tetrahedra are reinterpreted as fundamental quanta of the GFT field which are created or annihilated by $2$nd quantized field operators. The Fock vacuum, defined by
\be
\hat{\varphi}(g_I)|\emptyset\rangle=0,
\ee
corresponds to the no-space state which is devoid of any topological and quantum geometric information. By convention, it is normalized to $1$. The excitation of a GFT quantum over the Fock vacuum, expressed by
\be
|g_I\rangle=\hat{\varphi}^{\dagger}(g_I)|\emptyset\rangle,
\ee
corresponds then to the creation of a single open $4$-valent LQG spin network vertex. The field operators obey the Canonical Commutation Relations
\be
[\hat{\varphi}(g_I),\hat{\varphi}^{\dagger}(g'_I)]=\mathbb{1}_G(g_I,g'_I)~\textrm{and}~[\hat{\varphi}^{(\dagger)}(g_I),\hat{\varphi}^{(\dagger)}(g'_I)]=0,
\ee
where the delta distribution $\mathbb{1}_G(g_I,g'_I)=\int_G dh \prod_I\delta(g_I h g_{I}'^{-1})$ on $G^d/G$ is given in such a way as to be compatible with the right invariance of the fields, as defined in Eq. (\ref{Rightinvariance}). Using this technology, the construction of properly symmetrized $N$-particle states is possible which are needed to describe extended quantum $3$-geometries.

Quantum geometric observable data can be retrieved from such states by means of second-quantized Hermitian operators \cite{GFTLQG,GFTOperators}. For example, the vertex number operator is expressed by
\be 
\hat{N}=\int(dg)^d\hat{\varphi}^{\dagger}(g_I)\hat{\varphi}(g_I)
\ee
and the vertex volume operator is encoded by
\be
\hat{V}=\int(dg)^d\int(dg')^d\hat{\varphi}^{\dagger}(g_I)V(g_I,g'_I)\hat{\varphi}(g'_I)
\ee
where $V(g_I,g'_I)=\langle g_I|\hat{v}|g'_I\rangle$ is given in terms of the matrix elements of the first-quantized LQG volume operator $\hat{v}$ \cite{LQG,LQGdiscreteness}.

\subsection{Group Field Theory Condensate Cosmology}\label{gfcrecap}

The goal of the GFT condensate cosmology program is to give a description of cosmologically relevant geometries in terms of the above-given technology. More specifically, one tries to model homogeneous continuum $3$-geometries and their cosmological evolution by means of GFT condensate states and their effective dynamics.

To this aim, it is conjectured that a phase transition in a GFT system gives rise to a condensate phase which in turn corresponds to a nonperturbative vacuum of the specific model under consideration. This vacuum is described by a large number $N=\langle \hat{N}\rangle$ of bosonic GFT quanta which have relaxed into a common ground state that is asymptotically orthogonal to the Fock vacuum, i.e. in the limit where $N\to\infty$ (cf. \cite{BECs}). This feature makes them suitable to model spatially homogeneous (quantum) geometries, whose metric is supposed to be the same at every point of the space emerging from the condensate. We will also require that only almost flat building blocks are chosen so that in combination with a large enough constituent number $N$ a smooth continuum is well approximated. In addition, since condensate states are field coherent states they exhibit semiclassical properties. This is required such that a classical cosmological spacetime can emerge from the underlying quantum description \cite{GFC,GFCExample,GFCExample2,GFCOthers,GFCEmergentFriedmann,GFCEmergentFriedmann2,
GFClowspin,GFClowspinStaticEffInt,GFCEmergentFriedmann3,GFCReview}.

For the construction of suitable states, we follow the Bogoliubov approximation which is valid for ultracold, non- to weakly interacting and dilute real Bose condensates \cite{BECs,BogoliubovAnsatz}. When the ground state is macroscopically occupied, one can separate the field operator into 
\be
\hat{\varphi}(g_I)=\sigma(g_I)\mathbb{1}+\delta\hat{\varphi}(g_I)
\ee
where $\sigma$ denotes the condensate mean field and $\delta\hat{\varphi}$ takes into account the remaining noncondensate contributions. The expectation value of the field operator is in general nonzero, which indicates that the condensate state is in or close to a coherent state. 

A simple trial state which fulfills this ansatz is given by
\be
|\sigma\rangle=A~e^{\hat{\sigma}}|\emptyset\rangle,~~~\hat{\sigma}=\int(dg)^d\sigma(g_I)\hat{\varphi}^{\dagger}(g_I),
\ee
with normalization factor $A=e^{-\frac{1}{2}\int(dg)^d|\sigma(g_I)|^2}$. It is constructed from quantum tetrahedra which encode all the same quantum geometric information.\footnote{More complicated``molecule" states built from the tetrahedra could also be considered, as advocated in Refs. \cite{GFC,GFCReview}.} These are field coherent states since they are eigenstates of the field operator, i.e.,
\be
\hat{\varphi}(g_I)|\sigma\rangle=\sigma(g_I)|\sigma\rangle
\ee
implying that $\langle\hat{\varphi}(g_I)\rangle=\sigma(g_I)\neq 0$ holds.\footnote{To guarantee the invariance under local frame rotaions, one also requires $\sigma(g_I)$ to be invariant with respect to the left diagonal action of $G$.}

In order to obtain information about the dynamics of this state, at least in an approximate sense, one performs a saddle point approximation on the path integral, which is this time constructed with the above-introduced coherent states. The effective dynamics is then captured by
\be\label{GPGFC}
\frac{\delta S[\sigma,\bar{\sigma}]}{\delta\bar{\sigma}(g_I)}=\int(dg')^d\mathcal{K}(g_I,g'_I)\sigma(g'_I)+\frac{\delta\mathcal{V}}{\delta\bar{\sigma}(g_I)}=0,
\ee
with the action $S[\sigma,\bar{\sigma}]$ as in Eq. (\ref{GFTaction}), but now written in terms of the condensate field $\sigma$. This mean field equation is the analogue of the Gross-Pitaevskii (GP) equation for real Bose condensates and has definite classical features. It is interpreted as a quantum cosmology equation, however, like the GP equation, it has no direct probabilistic interpretation \cite{BECs}. As shown below, this does not form an obstacle to extract cosmological predictions from the theory. In particular, it allows us to retrieve Friedmann-like evolution equations from the effective dynamics of such GFT condensate states.

In a final step, the quantum dynamics of the mean field $\sigma$ is specified by choosing the kinetic kernel $\mathcal{K}$ and the interaction term in the action. Typically, $\mathcal{K}$ is given by
\be\label{kineticoperatorgeneral}
\mathcal{K}=\delta(g'_{I}g^{-1}_{I})\delta(\phi'-\phi)\biggl[-(\tau\partial_{\phi}^2+\sum_{I=1}^4\Delta_{g_I})+M^2\biggr],~~~\tau>0.
\ee
Some remarks are in order. First, a free massless scalar field $\phi$ was added to the action serving as a relational clock with respect to which we study the evolution of the condensate. This requires that the GFT field is now written as $\varphi=\varphi(g_I,\phi)$.\footnote{For more details about the introduction of the clock field into the GFT action we refer the reader to Ref. \cite{GFCEmergentFriedmann}.} Second, to guarantee that the partition function is bounded from below, the signs of the terms in $\mathcal{K}$ have to be chosen accordingly.\footnote{In this way, the functional $S$ resembles an Euclideanized action. This should be kept in mind when computing the equation of motion of the corresponding dynamical system as for example in Eq. (\ref{eomfree}), thus necessitating a Wick rotation of $\phi$ to $i\phi$ as in Ref. \cite{GFCExample2}.} Third, the Laplace-Beltrami operator on the group manifold is motivated by the renormalization group analysis of GFT models where it is shown to be indispensible in order to regulate the UV behavior of the theory \cite{GFTRGReview}. Finally, the ``mass term" is related to the GFT/spin foam correspondence, where it is shown that it corresponds to the spin foam edge weights \cite{GFT,GFTSF}.

Below we will be concerned with the analysis of GFT systems based on real-valued scalar fields, that means for which $\hat{\varphi}(g_I)=\hat{\varphi}^{\dagger}(g_I)$ holds. This simplifies some calculations and, as we will see, leads to some extent to a different phenomenology compared to complex-valued fields (cf. Refs. \cite{GFCEmergentFriedmann,GFCEmergentFriedmann2,GFCEmergentFriedmann3}).

The selection of a suitable interaction term $\mathcal{V}$ completes the choice of model. In the remainder of this article we will study the impact of combinatorially local interaction terms onto the dynamics of the system. From the point of view of mean field theory, it is not uncommon to study simplified types of interactions which gloss over the actual microscopic details. In a same manner, we speculate that combinatorial local interactions between the condensate constituents are only relevant in a continuum and large scale limit, where the true combinatorial nonlocality of the fundamental theory could be effectively hidden. Hence, by adopting a rather phenomenological perspective, we hope to clarify the map between the microscopic and macroscopic regimes of the theory. In the end, rigorous Renormalization Group arguments will have the last word on whether local interactions can at all be derived from the fundamental theory \cite{FRG,MM,TGFTFRG,FRGcurvedspace,TMRG,GFTRGReview}.  

\section{Static case of an isotropic and free GFT condensate configuration}\label{sectionstatic}

In this section we focus on ``static" mean fields, i.e. $\sigma(g_I,\phi)=\sigma(g_I)$. Within  subsection \ref{recapstaticfree} we recapitulate some results of Ref. \cite{GFClowspinStaticEffInt} obtained for a static and free GFT condensate in an isotropic restriction. In subsection \ref{ncftstaticfree} we present the noncommutative Fourier transform for such condensate wave functions from which it is in principle possible to retrieve the $3$-metric. 

We start by neglecting all interactions, i.e., $\mathcal{V}=0$. Then Eq. (\ref{GPGFC}) together with Eq. (\ref{kineticoperatorgeneral}) yield
\be\label{StaticFreeGP}
\biggl[-\sum_{I=1}^4\Delta_{g_I}+M^2\biggr]\sigma(g_I)=0.
\ee
To solve this equation of motion, we can introduce coordinates on the $\textrm{SU}(2)$ group manifold and use the left and right invariance properties of the GFT field. Hence, $\sigma(g_I)$ lives on the domain space $\textrm{SU}(2)\backslash\textrm{SU}(2)^4/\textrm{SU}(2)$ which can be parametrized by six independent coordinates.

To illustrate this, assume that the connection in the holonomy $g=\mathcal{P}e^{i\int_e A}$ is approximately constant along the link $e$ with length $\ell_0$ in the $x$-direction, which yields $g\approx e^{i\ell_0 A_x}$. Using the polar decomposition, one has 
\be
g=\cos(\ell_0||\vec{A}_x||)\mathbbm{1}+i\vec{\sigma}\frac{\vec{A}_x}{||\vec{A}_x||}\sin(\ell_0||\vec{A}_x||),
\ee 
with the $\mathfrak{su(2)}$-connection $A_x=\vec{A}_x\cdot\vec{\sigma}$ and the Pauli matrices $\{\sigma_i\}_{i=1...3}$. In a second step, one introduces the coordinates $(\pi_0,...,\pi_3)$ on the group manifold together with $\pi_0^2+...+\pi_3^2=1$ specifying an embedding of $\mathrm{SU}(2)\cong S^3$ into $\mathbb{R}^4$. Because of the isomorphism $\textrm{SO}(3)\cong \textrm{SU}(2)/\mathbb{Z}_2$,  choosing the sign in $\pi_0=\pm\sqrt{1-\vec{\pi}^2}$ will correspond to the parametrization of just one hemisphere of the $3$-sphere. Finally, by identifying 
\be\label{coordinatesongroup}
\vec{\pi}=\frac{\vec{A}_x}{||\vec{A}_x||}\sin(\ell_0||\vec{A}_x||),
\ee
the holonomies are parametrized as
\be
g(\vec{\pi})=\sqrt{1-\vec{\pi}^2}\mathbbm{1}+i\vec{\sigma}\cdot\vec{\pi},
\ee
with $||\vec{\pi}||\leq 1$. At $||\vec{\pi}||=0$ we reach the pole of the hemisphere and at $||\vec{\pi}||=1$ the equator. 

Using the Lie derivative on the group manifold acting on a function $f$, one has for the Lie algebra elements
\be
\vec{B}f(g)\equiv i\frac{d}{dt}f(e^{\frac{i}{2}\vec{\sigma}t}g)|_{t=0}=\frac{i}{2}\biggl[\sqrt{1-\vec{\pi}^2}\vec{\nabla}+\vec{\pi}\times\vec{\nabla}\biggr]f.
\ee
With this the Laplace-Beltrami operator $\vec{B}^2=-\Delta_{g}$ in terms of the coordinates $\vec{\pi}$ on $\mathrm{SU}(2)$ is given by
\be 
-\Delta_gf(g)=-[(\delta^{ij}-\pi^i\pi^j)\partial_i\partial_j-3\pi^i\partial_i]f(\vec{\pi}).
\ee 
This applies to all group elements $g_I$, with $I=1,...,d$, dressing the spin network vertex dual to the quantum tetrahedron. Thus, the Laplacian part in the equation of motion Eq. (\ref{StaticFreeGP}) is given by
\be\label{Laplacefor4} 
-\sum_{I}\Delta_{g_{I}}=\sum_I\vec{B}_I\cdot\vec{B}_I.
\ee
Using the invariance of the mean field under the right diagonal action of $G$, which corresponds to the closure condition for the fluxes
\be
\sum_{I}\vec{B}_I=0,
\ee
as detailed in Appendix \ref{AppendixA}, Eq. (\ref{Laplacefor4}) can be rewritten as
\be\label{kinetictermLaplacian}
-\sum_{I}\Delta_{g_{I}}=2~\biggl(\sum_{i=1}^3\vec{B}_i\cdot\vec{B}_i+\sum_{i\neq j}\vec{B}_i\cdot\vec{B}_j\biggr).
\ee

In the most general case, the gauge invariant $\sigma(g_I)$ is defined on a domain space parametrized by six invariant coordinates $\pi_{ij}=\vec{\pi}_{i}\cdot\vec{\pi}_j$, with $i,j=1,2,3$ and $0\leq|\pi_{ij}|\leq 1$. Hence, when using the above considerations, the action of Eq. (\ref{kinetictermLaplacian}) in the equation of motion Eq. (\ref{StaticFreeGP}) will give rise to a rather complicated partial differential equation (cf. Ref. \cite{GFCExample}). In the following subsections we will study specific solutions to the latter.

\subsection{Static case of an isotropic and free GFT condensate configuration}\label{recapstaticfree}

A strategy to extract solutions, as outlined in Refs. \cite{GFCExample,GFCExample2}, is to impose in a first step a symmetry reduction by considering functions $\sigma$ which only depend on the diagonal components $\pi_{ii}$. In this way, Eq. (\ref{kinetictermLaplacian}) acting on $\sigma(\pi_{ii})$ leads to
\begin{multline}\label{laplacefirstlevelreduction}
-\sum_I\Delta_{g_{I}}\sigma(\pi_{ii})~=~\\-\biggl[\sum_i 8\pi_{ii}(1-\pi_{ii})\frac{\partial^2}{\partial\pi_{ii}^2}+ 4(3-4\pi_{ii})\frac{\partial}{\partial\pi_{ii}}\\+~4\sum_{i\neq j}\sqrt{1-\pi_{ii}}\sqrt{1-\pi_{jj}}\pi_{(ij)}\frac{\partial^2}{\partial\pi_{ii}\partial\pi_{jj}}\biggr]\sigma(\pi_{ii}).
\end{multline}
The second sum stems from the corresponding one in Eq. (\ref{kinetictermLaplacian}). To find particular solutions, it is advantageous to get rid of this term in order to decouple the terms in the different $\pi_{ii}$. To this purpose, one can employ a summation ansatz
\be
\sigma(\pi_{ii})=\sigma(\pi_{11})+...+\sigma(\pi_{33}).
\ee
Finally, assuming all $\pi_{ii}$ to be equal to a variable $p$, Eq. (\ref{StaticFreeGP}) can be rewritten as
\be\label{masterequation}
-\biggl[2p(1-p)\frac{d^2}{dp^2}+(3-4p)\frac{d}{dp}\biggr]\sigma(p)+\mu\sigma(p)=0,
\ee
with $\mu\equiv\frac{M^2}{12}$ and $p\in[0,1]$ for which analytic general solutions can be found \cite{GFCExample,GFCExample2}.

In retrospection, it is sensible to refer to such a reduction, i.e. to just one variable $p$, as an isotropization. This can be seen when substituting 
\be
p=\pi_{ii}\equiv\sin^2(\psi)
\ee
into Eq. (\ref{masterequation}) which leads to
\be\label{masterequationangle}
-[\frac{d^2}{d\psi^2}+2\cot(\psi)\frac{d}{d\psi}]\sigma(\psi)+2\mu\sigma(\psi)=0,~~~\psi\in[0,\frac{\pi}{2}].
\ee
Comparing this expression to the Laplacian on one hemisphere of $S^3$ which acts on a function $\sigma(\phi,\theta,\psi)$, one has
\be\label{LaplacianHemisphere}
-\Delta \sigma(\phi,\theta,\psi)=-\frac{1}{\sin^2(\psi)}\biggl[\frac{\partial}{\partial\psi}(\sin^2(\psi)\frac{\partial}{\partial\psi}\sigma)+\Delta_{S^2}\sigma\biggr],
\ee
with $\phi\in[0,2\pi]$, $\theta\in[0,\pi]$ and $\psi\in[0,\frac{\pi}{2}]$. The function $\sigma$ is called hyperspherically symmetric, isotropic or zonal if it is independent of $\phi$ and $\theta$ \cite{AnalysisonLieGroups}. These are spherically symmetric eigenfunctions of $-\Delta_{S^2}$ for which Eq. (\ref{masterequationangle}) is just equal to Eq. (\ref{LaplacianHemisphere}). We can interpret this symmetry reduction as an explicit restriction of the rather general class of condensates to a particular representative with a clearer geometric interpretation. In addition, we note that this reduction should not be confused with the symmetry reduction employed in Wheeler-DeWitt quantum cosmology \cite{WdWQC} or LQC \cite{LQC}, since it is applied after quantization onto the quantum state and not beforehand.

To specify the spectrum of the differential equation, one introduces boundary conditions inferred from physical assumptions. Since we are looking for solutions that admit an interpretation in terms of smooth metric $3$-geometries the curvature per condensate constituent is supposed to be small. In the group-representation, this so-called near-flatness condition concretely translates into demanding that the character of the group elements decorating the quantum tetrahedra are close to $\chi(\mathds{1}_{j})=2j+1$ according to \cite{GFC, GFCReview}. 

For the mean field this means that the probability density is concentrated around small values of the connection or its curvature. In our symmetry reduced scenario this is realized for $\sigma(\psi)$ if the probability density $|\sigma(\psi)|^2$ has its global maximum around small values of the variable $\psi$ and we require that it tends to zero at the equator traced out at $\psi=\frac{\pi}{2}$ after which it remains zero until $\pi$ is reached.\footnote{More generally, one could require the probability density to be small against its value at $\psi=0$. However, this would only have a marginal impact onto the subsequent discussion. Instead it would be important to check if such a behavior of the probability density can follow by means of a dynamical principle.} This translates into a Dirichlet boundary condition on the equator,
\be
\sigma(\psi)|_{\psi=\frac{\pi}{2}}=0.
\ee
The eigensolutions obeying this boundary condition are then given by
\be\label{solutionsangle}
\sigma_j(\psi)=\frac{\sin((2j+1)\psi)}{\sin(\psi)},~~~\psi\in[0,\frac{\pi}{2}]
\ee
with $j\in \frac{2\mathbb{N}_0+1}{2}$ corresponding to the eigenvalues $\mu=-2j(j+1)$. On the interval $[0,\frac{\pi}{2}]$ these solutions are exactly equal to those hyperspherically symmetric eigenfunctions of the Laplacian on $S^3$ which are zero on the equator. Furthermore, observe that these would just be the characters $\chi_j(\psi)$ of the respective representation for $j$ if $\sigma_j(\psi)$ was not zero from $\frac{\pi}{2}$ onwards. With this it is easy to see that our solutions obey the near-flatness condition, since $\lim_{\psi\to 0}\sigma_j(\psi)=2j+1$.\footnote{It can be justified to relate the variable $p$ to the field strength by considering a plaquette $\Box$ in one face of a tetrahedron together with the well-known expression
\be\nonumber
F^k_{ab}(A)=\frac{1}{\mathrm{Tr}(\tau^k\tau^k)}\lim_{\textrm{Area}_{\Box}\to 0}\mathrm{Tr}_j\biggl(\tau^k\frac{\mathrm{hol}_{\Box_{ij}}(A)-1}{\textrm{Area}_{\Box}}\biggr)\delta_a^i\delta^j_b,
\ee
where $a,b\in\{1,2\}$. For the $\mathfrak{su}(2)$-algebra elements $\tau_k=-\frac{i}{2}\sigma_k$ the relation $\mathrm{Tr}(\tau^k\tau^k)=-\frac{1}{3}j(j+1)(2j+1)$ holds \cite{LQC}. Hence, one obtains $F^k\sim \sin^2(\psi)=p$.}\footnote{From a heuristic point of view, using the first-order formalism for gravity, it is possible to argue that a small field strength implies a small $3$-curvature $R$. Hence, a concentration of the probability density around small $\psi$ directly corresponds to a concentration around small curvature values.}

In this representation the expectation value of the number operator yields
\be
N=\int_0^{\frac{\pi}{2}}d\psi \sin^2(\psi)~|\sigma(\psi)|^2.
\ee

In view of the further geometric interpretation of these solutions, one can transform these solutions to the spin-representation which facilitates most directly the extraction of information about the LQG volume and area operators. To this aim, we use that because of the left- and right-invariance of $\sigma(g_I)$, the mean field is in particular a central function on the domain space, i.e., $\sigma(h g_I h^{-1})=\sigma(g_I)$ for all $h\in \mathrm{SU}(2)$. This holds of course for the isotropic function $\sigma(\psi)$, too. Remarkably, the notion of isotropy coincides in this case with the notion of centrality \cite{AnalysisonLieGroups}. Using the Fourier series of a central function on $\mathrm{SU}(2)$, the Fourier series of the mean field in the angle parametrization is expressed by
\be 
\sigma_j(\psi)=\sum_{m\in\mathbb{N}_0/2}(2m+1)~\chi_m(\psi)~\sigma_{j;m},
\ee
with the ``plane waves" given by the characters 
\be
\chi_m(\psi)=\frac{\sin((2m+1)\psi)}{\sin(\psi)}.
\ee 
One then obtains the Fourier coefficients by
\be
\sigma_{j;m}=\frac{2}{\pi}\frac{1}{2m+1}\int_{0}^{\frac{\pi}{2}}d\psi \sin^2(\psi)~\chi_m(\psi)~\sigma_j(\psi),
\ee
with $m\in\frac{\mathbb{N}_0}{2}$. Notice that in principle this applies to more general solutions $\sigma(\psi)$, too.\footnote{It should be remarked that the double index in $\sigma_{j;m}$ accounts for the fact that $\sigma_j(\psi)$ is only nonzero on the interval $[0,\frac{\pi}{2})$ and vanishes on $[\frac{\pi}{2},\pi]$.} 

\begin{figure}[ht]
	\centering
  \includegraphics[width=0.4\textwidth]{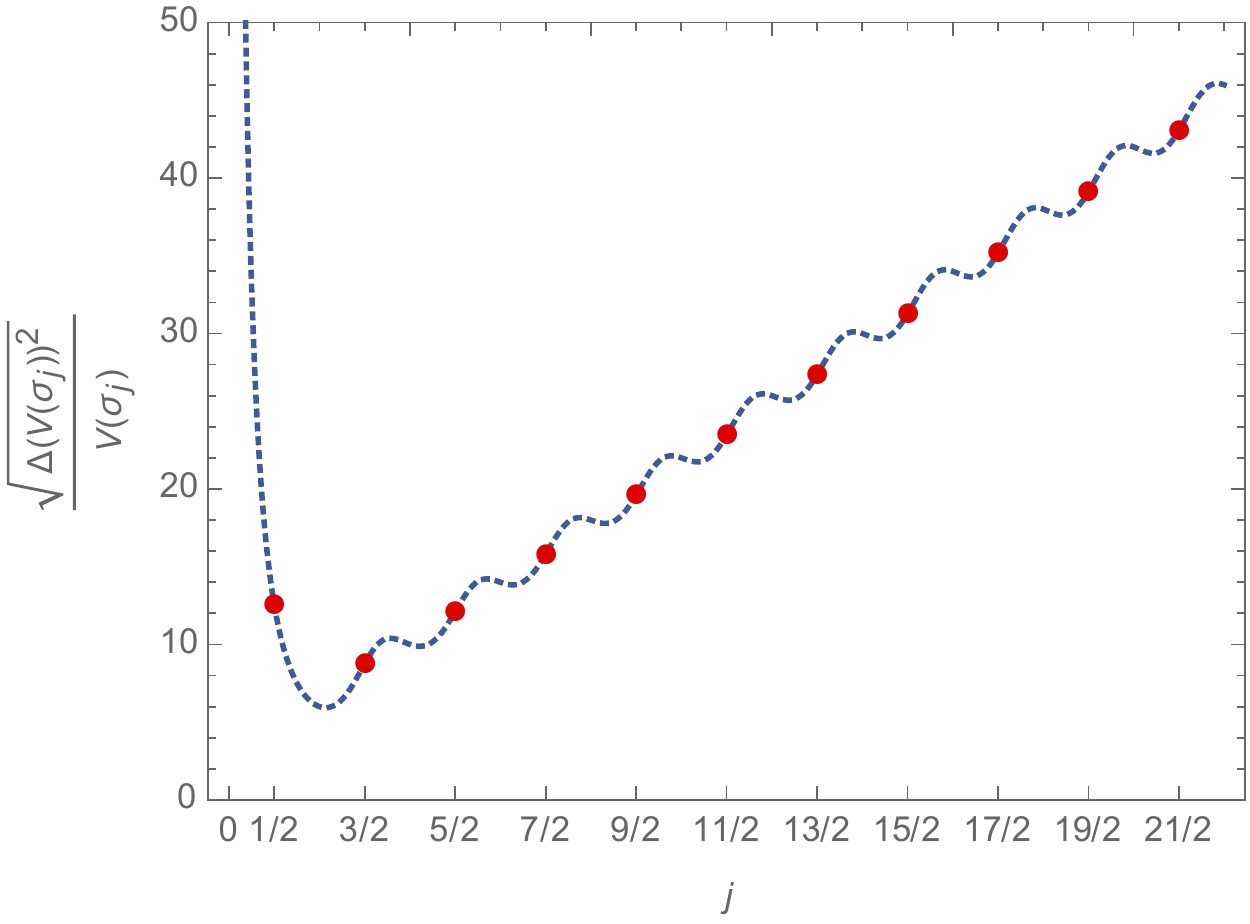}
\caption{Relative standard deviation of the volume operator over $j$.}
\label{RRMSVolumen}
\end{figure}

Having this at hand, one can calculate the expectation value, e.g., of the LQG volume operator with respect to the mean field in the spin-representation. It is given by
\be
V_j=V_0\sum_{m\in\mathbb{N}_0/2} |\sigma_{j;m}|^2 V_m~~\textrm{with}~~V_m\sim m^{3/2}
\ee
and $V_0\sim\ell_p^3$. This object is central to the geometric interpretation of solutions to the equation of motion. For instance, using the expectation value of the volume operator one can compute the relative standard deviation of the volume contributions $V_j$ for the solutions Eq. (\ref{solutionsangle}) which is illustrated in Fig. \ref{RRMSVolumen}. This shows that the relative uncertainty of the $V_j$s increases linearly in $j$. In subsection \ref{relationalevolutioniso} we will reconsider this quantity in the context of the relational evolution of the total volume. We will then show that the relative uncertainty vanishes at late times which indicates the classicalization of the quantum geometry emerging from the condensate state. 

\subsection{Static case of an isotropic and free GFT condensate configuration: Noncommutative Fourier transform}\label{ncftstaticfree}

The present section makes use of the notion of the noncommutative Fourier transform $\mathcal{F}$ on $\mathrm{SU}(2)$ or $\mathrm{SO}(3)$, respectively. $\mathcal{F}$ defines an isometric mapping between the space $L^2(\mathrm{SO}(3),d\mu_{H})$ with Haar measure $d\mu_{H}$ and the space $L^2_{\star}(\mathbb{R}^3,d\mu)$ of functions on $\mathfrak{su}(2)\sim \mathbb{R}^3$ with a noncommutative $\star$-product and standard Lebesgue measure $d\mu$, as expounded in greater detail in \cite{NCFT}. This transform allows us to shift in between the group and the dual flux representation of the mean field $\sigma$, which we review in Appendix \ref{AppendixA}. From the momentum space representation of the condensate field it is in principle possible to reconstruct the metric at a given point in one of the quantum tetrahedra constituting the condensate \cite{GFC,GFCReview}. 

In the following, we make use of the coordinates introduced in Section \ref{sectionstatic} to parametrize the group manifold $\mathrm{SU}(2)$. In particular, since we are working on $\mathrm{SO}(3)$ which can be identified to the upper hemisphere of $\mathrm{SU}(2)\cong S^3$, we adapt our coordinates to $\vec{\pi}=\sin(\psi)\vec{n}$ with $\psi\in[0,\frac{\pi}{2}]$ and $\vec{n}\in S^2$. In this parametrization the Haar measure is given by $dg=\frac{1}{\pi}\sin^2(\psi)~d\psi~d^2\vec{n}$, where $d^2\vec{n}=\sin(\theta)~d\phi~d\theta$ is the normalized measure on the unit $2$-sphere. With this one can recast the transform given by Eq. (\ref{ncFT}) into a standard $\mathbb{R}^3$-Fourier transform, leading to the integral formula
\be
\mathcal{F}[\sigma](x)=\frac{1}{\pi}\int_{||\vec{\pi}||\leq 1}\frac{d^3\vec{\pi}}{\sqrt{1-\vec{\pi}^2}}~\sigma(g(\vec{\pi}))~e^{i\vec{\pi}\cdot\vec{B}}.
\ee
Taking advantage of the symmetry of the problem in the isotropic restriction, together with $x\equiv ||\vec{B}||$ and $||\vec{n}||=1$, one finds that $\vec{\pi}\cdot\vec{B}= x \sin(\psi)\cos(\theta)$. In a next step one integrates over $\theta$ which leads with $p\equiv\vec{\pi}^2$ to the analog of the Fourier-Bessel transformation given by
\be
\hat{\sigma}(x)=2\int_0^1 \frac{dp}{\sqrt{1-p}}\frac{\sin(\sqrt{p}~x)}{x}~\sigma(\sqrt{p}).
\ee
In a final step, we use $p\equiv \sin^2(\psi)$ to arrive at an expression in terms of the angles $\psi$ given by
\be
\hat{\sigma}(x)=4\int_{0}^{\frac{\pi}{2}}d\psi~\sin(\psi)\frac{\sin(\sin(\psi)x)}{x}~\sigma(\psi).
\ee

\begin{figure}[ht]
	\centering
  \includegraphics[width=0.4\textwidth]{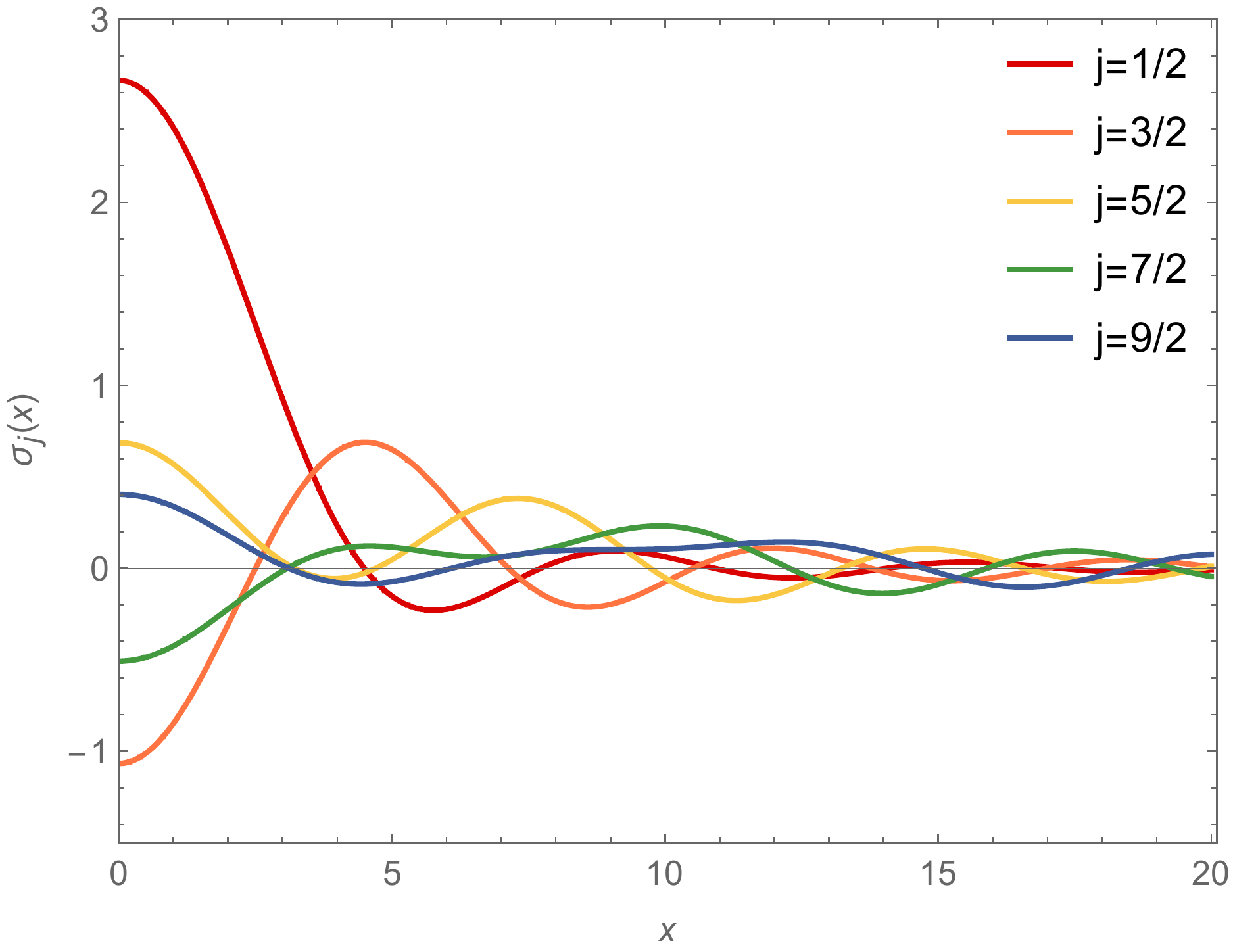}
\caption{Noncommutative Fourier transform of the free isotropic solution $\sigma_j(\psi)$.}
\label{NCFTexample}
\end{figure}

Using this, we can compute the noncommutative Fourier transform $\hat{\sigma}_j(x)$ of the solution to the free equation given by Eq. (\ref{solutionsangle}). The results for different $j$ are illustrated in Fig. \ref{NCFTexample}.

Using Appendix \ref{AppendixA} together with the flux-information encoded by the noncommutative Fourier transform $\hat{\sigma}_j(x)$, one can relate the geometric content of the quantum tetrahedra to a metric for this (or any other) model. The reconstruction of the metric $g_{ij}$, as recapitulated by Eq. (\ref{reconstructedmetric}), however, is somewhat obstructed, e.g., due to ordering ambiguities in the corresponding operator version $\hat{g}_{ij}$ stemming from the noncommutativity of the fluxes, as noticed in Ref. \cite{GFCReview}.

\section{Relational evolution of an isotropic GFT condensate configuration}\label{relationalevolutioniso}

In the subsequent section we investigate aspects of the relational evolution of free and effectively interacting GFT condensates in the above introduced isotropic restriction. In subsection \ref{lowspinisotropic} we show that such a condensate quickly settles into a low-spin configuration in the free case, which is similar to the results obtained in Ref. \cite{GFClowspin}. Using this, in subsection \ref{emergentfriedmanndynamics} it is demonstrated how the classical Friedmann dynamics then emerge in the semiclassical limit, as in Ref. \cite{GFCEmergentFriedmann}. Following the techniques developed in Ref. \cite{GFCEmergentFriedmann3}, we show that the expansion of the emergent space is accelerated, though not strong enough to supplant inflation. 

In subsection \ref{isotropicthomasfermi} we  study the influence of effective interactions onto the evolution of the condensate,  by considering firstly the equivalent to the Thomas-Fermi approximation for real BECs \cite{BECs,NumericalrealBECs} which allows us to conduct a fixed point analysis of the remaining dynamical system. Then, in subsection \ref{isotropicacceleratedexpansion} we investigate the acceleration behavior of such effectively interacting condensates. There we show, by closely following Ref. \cite{GFCEmergentFriedmann3}, that for two interaction terms one can in principle find a sufficiently strong acceleration to replace the inflationary paradigm, however, coming at the cost of fine-tuning their coupling constants. Finally, in subsection \ref{isotropiclinearizedsystem} we investigate the formal stability properties of the full dynamical system including the Laplace-Beltrami and interaction term. This allows us to determine which modes, depending on the ``mass parameter", are (in the language of dynamical systems) stable or unstable.

\subsection{Relational evolution of an isotropic and free GFT condensate configuration: Emergence of a low-spin phase}\label{lowspinisotropic}

We employ the idea used in Ref. \cite{GFClowspin}  regarding the dynamical relaxation of a GFT condensate system into a low-spin phase. However, in contrast to the strategy followed in Ref. \cite{GFClowspin}, we will not work from the beginning in the spin-representation and will not make use of the notion of isotropy introduced in Ref. \cite{GFCEmergentFriedmann}. Instead, we work in the coordinate-representation, which together with the symmetry reductions recapitulated above, will lead us to a qualitatively equivalent result. In addition, we illustrate the values of the parameters for which exponentially dominating low-spin configurations can be found. 

To this aim, we start with the equation of motion for the free case,
\be\label{eomfree}
\biggl[(\tau\partial^2_{\phi}-\sum_{I=1}^4\Delta_{g_{I}})+M^2\biggr]\sigma(g_I,\phi)=0,
\ee
which takes account of the evolution with respect to the relational clock $\phi$. With the above discussed isotropic restriction this yields the partial differential equation (PDE)
\be
\biggl[(2\tau\partial^2_{\phi}-\frac{d^2}{d\psi^2}-2\cot(\psi)\frac{d}{d\psi})+2\mu\biggr]\sigma(\psi,\phi)=0,~~\psi\in[0,\frac{\pi}{2}].
\ee
General solutions to this PDE can be obtained by employing a separation ansatz which is allowed because the terms in $\phi$ and $\psi$ completely decouple. With $\sigma(\psi,\phi)=\xi(\psi)T(\phi)$ one yields,
\be
2\tau\frac{\partial_{\phi}^2T(\phi)}{T(\phi)}=\frac{(\frac{d^2}{d\psi^2}+2\cot(\psi)\frac{d}{d\psi}-2\mu)\xi(\psi)}{\xi(\psi)}\equiv\omega=\mathrm{const}.
\ee
The general solution to
\be
\partial_{\phi}^2 T(\phi)=\frac{\omega}{2\tau}~ T(\phi)~,
\ee
is given by
\be\label{zeitanteil}
T(\phi)=\biggl(\mathrm{a}_1~e^{\sqrt{\frac{\omega}{2\tau}}\phi}+\mathrm{a}_2~e^{-\sqrt{\frac{\omega}{2\tau}}\phi}\biggr),
\ee
with the constants $\mathrm{a}_1,\mathrm{a}_2\in\mathbb{C}$.

The spectrum of the differential equation depending solely on $\psi$ is concretized by imposing boundary conditions. In accordance with the near-flatness condition one chooses the Dirichlet boundary condition $\xi(\psi=\frac{\pi}{2})=0$ so that the eigensolutions of the $\psi$-part are  given by
\be
\xi_{j}(\psi)=\frac{\sin((2j+1)\psi)}{\sin(\psi)},~~\psi\in[0,\frac{\pi}{2}]
\ee
with eigenvalues $\omega+2\mu=-4j(j+1)$ and $j\in\frac{2\mathbb{N}_0+1}{2}$.\footnote{Notice that $\xi_j(\psi)$ is zero on the interval $[\frac{\pi}{2},\pi]$ which leads to a double index in $\sigma_{j;m}(\phi)$ from Eq. (\ref{expectationvaluevolume}) onward as remarked for the static case in subsection \ref{recapstaticfree}.} 

In the following, we want to study the evolution of the solutions $\sigma_j(\psi,\phi)=\xi_{j}(\psi)T_j(\phi)$ with respect to the relational clock $\phi$ with particular regard to their behavior for $\phi\to 0$ and $\phi\to\pm\infty$. Using the form of the spectrum, we can substitute $\omega$ into Eq.~(\ref{zeitanteil}) such that
\be\label{timefunction}
T_j(\phi)=\biggl(\mathrm{a}_1~e^{\sqrt{\frac{\mu+2j(j+1)}{-\tau}}\phi}+\mathrm{a}_2~e^{-\sqrt{\frac{\mu+2j(j+1)}{-\tau}}\phi}\biggr)
\ee
whose behavior critically depends on the sign of $\frac{\mu+2j(j+1)}{-\tau}$ with $\mu<0$ and $\tau>0$. For two possible initial conditions $T_j(0)=0$ and $T'_j(0)=0$ the solutions are
\be
T_j(\phi)=2\mathrm{a}_1\sinh\biggl(\sqrt{\frac{\mu+2j(j+1)}{-\tau}}\phi\biggr)
\ee
and
\be
T_j(\phi)=2\mathrm{a}_1\cosh\biggl(\sqrt{\frac{\mu+2j(j+1)}{-\tau}}\phi\biggr),
\ee
respectively. Importantly, only in the latter case  $\lim_{\phi\to 0} T_j(\phi)\neq 0$ holds leading to a nonvanishing volume, as we shall see below. Hence, for such solutions the singularity problem is avoided.\footnote{In principle, constructing solutions which vanish at $\phi=0$ seems to be arbitrary. Instead, one could use that Eq. (\ref{timefunction}) vanishes for $\phi=\sqrt{\frac{2\tau}{\omega}}\frac{\log(-\mathrm{a}_1/\mathrm{a}_2)}{2}$ but this would not change the essence of our argument.}\footnote{We want to remark that the resolution of the singularity problem for real-valued GFT fields thus depends on the choice of initial conditions. This is clearly different to the situation found for complex-valued GFT fields in Refs. \cite{GFCEmergentFriedmann, GFCEmergentFriedmann2}. The occurrence of a bounce in these works is deeply rooted in the global $\mathrm{U}(1)$-symmetry of the complex-valued GFT field to which the conserved charge $Q$ can be associated, preventing the field from becoming zero for all values of the relational clock $\phi$ as long as $Q$ is nonzero. This mechanism does not depend on the initial conditions assigned to the mean field.}

\begin{figure}[ht]
	\centering
  \includegraphics[width=0.4\textwidth]{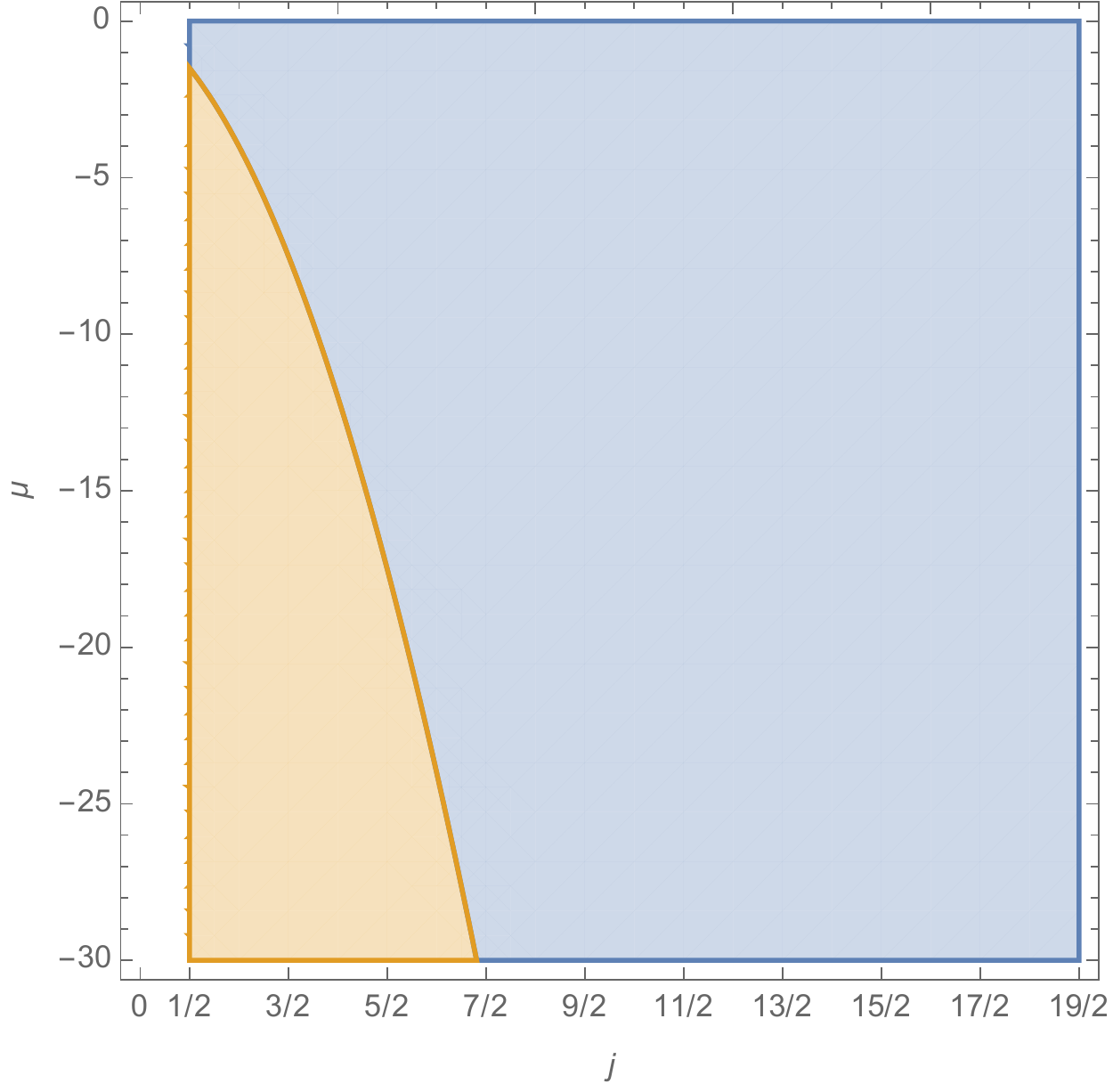}
	\caption{Two sectors representing exponentially growing (orange) and oscillating (blue) solutions.}
	\label{figparameterlowspin}
\end{figure}

Solutions for which
\be
\frac{1}{2}\leq j < -\frac{1}{2}+\frac{1}{2}\sqrt{1-2\mu}~~\mathrm{with}~~\mu< -\frac{3}{2} 
\ee
grow exponentially for $\phi\to\pm\infty$ whereas solutions with
\be
j\geq -\frac{1}{2}+\frac{1}{2}\sqrt{1-2\mu}~~\mathrm{with}~~\mu\leq-\frac{3}{2}
\ee
or
\be
j\geq \frac{1}{2}~~\mathrm{with}~~\mu> -\frac{3}{2}
\ee
display an oscillating behavior. This is illustrated by means of the differently colorized sectors in Fig.~\ref{figparameterlowspin}. Using this, we see that the condensate will quickly be dominated by the lowest representation  $j=\frac{1}{2}$ and all others are exponentially suppressed.

To extract further information about the behavior of the total volume of a general solution
\be
\sigma(\psi,\phi)=\sum_{j\in\frac{2\mathbb{N}_0+1}{2}}\xi_j(\psi)T_j(\phi),
\ee
we have to compute the Fourier components $\xi_{j;m}$ of $\xi_j(\psi)$ as in subsection \ref{recapstaticfree}. Using this, the expectation value of the volume operator with respect to the mean field $\sigma$ is decomposed as
\be\label{expectationvaluevolume}
\langle \hat{V}(\phi)\rangle\equiv V(\phi)=V_0\sum_{j\in\frac{2\mathbb{N}_0+1}{2}}\sum_{m\in\mathbb{N}_0/2} |\sigma_{j;m}(\phi)|^2 ~V_m
\ee
with $V_m\sim m^{3/2}$ and $V_0~\sim \ell_0^3$. 

From the relational evolution of the volume we deduce that only when $\lim_{\phi\to 0}T_j(\phi)\neq 0$ holds, the volume will be nonvanishing which indicates that singularity problem can be avoided.

At large $\phi$, the dominant contribution to the total volume is then given by
\be
\lim_{\phi\to\pm\infty}V(\phi)=V_0~V_{1/2}~|\xi_{\frac{1}{2};\frac{1}{2}}|^2~|\mathrm{a}_{1,2}|^2~e^{\pm 2\sqrt{\frac{\mu+\frac{3}{2}}{-\tau}}\phi}.
\ee
Bearing the slight differences in the sign conventions for $\mu$ and $\tau$ compared to Ref. \cite{GFClowspin} in mind, this result is, up to a factor of $1/2$ in the argument of the exponential function, just identical to the one obtained there. This tiny difference seems to reflect the differences in the imposition of the notion of isotropy and it is rather interesting that both also quantitatively lead to nearly the same behavior at late times. In addition, we want to remark that one also here obtains the direct proportionality between the asymptotic total number of quanta $N$ and the volume $V$ which in the context of LQC is needed for the so-called improved dynamics scheme (cf. Refs. \cite{LQCimproveddynamics, LQC}), as pointed out earlier in Ref.~\cite{GFClowspin}.

A last remark is in order. When computing the relative standard deviation or uncertainty of the volume operator with respect to the coherent state $\sigma$, given by
\be
\epsilon=\frac{\sqrt{\langle \hat{V}^2\rangle_{\sigma}-\langle\hat{V}\rangle_{\sigma}^2}}{\langle\hat{V}\rangle_{\sigma}},
\ee
one can show that it is dominated by the contributions of the $1/2$-mode at late times. In particular, one finds that
\be
\lim_{\phi\to\pm\infty}\epsilon~\sim~|\xi_{\frac{1}{2};\frac{1}{2}}|^{-1}~e^{\mp \sqrt{\frac{\mu+\frac{3}{2}}{-\tau}}\phi}\to 0,
\ee
which shows that measured values of the volume are tightly clustered around its mean. This suggests that the quantum geometry classicalizes at late times if the condensate system relaxes into a low-spin configuration.

\subsection{Relational evolution of an isotropic and free GFT condensate configuration: Emergent Friedmann dynamics and accelerated expansion}\label{emergentfriedmanndynamics}

In the following, we make use of the fact that the equation of motion Eq. (\ref{GPGFC}) together with the kinetic kernel Eq. (\ref{kineticoperatorgeneral}) in the free case lead to the conserved quantity,
\be\label{GFTEnergy}
E=\frac{1}{2}\int(dg)^4\biggl[\tau(\sigma')^2-\sigma\Delta\sigma+M^2\sigma^2\biggr],
\ee
to which one refers as the ``GFT energy" \cite{GFCEmergentFriedmann}. From a fundamental point of view, its physical meaning has yet to be clarified. 

When working in the above-introduced isotropic restriction with $\mu\equiv\frac{M^2}{12}$ and employing the Fourier decomposition of $\sigma(\psi,\phi)$ this expression can be rewritten into
\begin{equation}\label{GFTEnergydecomposition}
E
=\sum_{j\in\frac{2\mathbb{N}_0+1}{2}}\sum_{m\in\mathbb{N}_0/2}E_{j;m}
\end{equation}
where
\be\label{GFTEnergyjm}
E_{j;m}=\frac{1}{2}\biggl[\tau \sigma'^2_{j;m}(\phi)+[2(m(m+1))+\mu]\sigma^2_{j;m}(\phi)\biggr]~.
\ee
Using the introduction of the relational clock, the first Friedmann equation can be written as
\be
H^2=\Biggl(\frac{V'}{3 V}\biggr)^2\Biggl(\frac{d\phi}{dt}\Biggr)^2
\ee
where $t$ denotes the proper time. Together with the expectation value of the volume operator Eq. (\ref{expectationvaluevolume}) one obtains 
\begin{multline}
\Biggl(\frac{V'}{3 V}\biggr)^2=\Biggl(\frac{2\sum\limits_{j,m}V_m\sigma_{j;m}\sqrt{\frac{(2E_{j;m}-[2m(m+1)+\mu])\sigma^2_{j;m}}{\tau}}}{3\sum\limits_{j,m}V_m\sigma^2_{j;m}}\Biggr)^2
\end{multline}
where Eq. (\ref{GFTEnergyjm}) was used and the argument in $\sigma_{j;m}(\phi)$ is suppressed from hereon. Given that in the broken phase one has $\mu<0$, in the semiclassical limit, where $\sigma^2_{j;m}$ is large (cf. Ref. \cite{GFCEmergentFriedmann}), one finds
\be
\Biggl(\frac{V'}{3V}\biggr)^2\approx\Biggl(\frac{2\sum\limits_{j,m}V_m\sigma_{j;m}\sqrt{\frac{[|\mu|-2m(m+1)]\sigma^2_{j;m}}{\tau}}}{3\sum\limits_{j,m}V_m\sigma^2_{j;m}}\Biggr)^2.
\ee
Finally, exploiting the fact that the mean field quickly settles dynamically into the $j=1/2$ configuration, as obtained in the previous subsection, with the identification\footnote{More precisely, in Ref. \cite{GFCEmergentFriedmann2} it was shown that this identification holds asymptotically when the value of the relational clock grows.} $3\pi G=\frac{1}{\tau}[|\mu|-3/2]$ one recovers the classical Friedmann equations
\be
\Biggl(\frac{V'}{3V}\biggr)^2=\frac{4\pi G}{3}
\ee
and
\be
\frac{V''}{V}=12\pi G,
\ee
both given in terms of the relational clock $\phi$. We note that here we neglect the quantum gravity corrections to these equations stemming from the term proportional to $E_{\frac{1}{2};\frac{1}{2}}$ which is of the order of $\sqrt{\hbar G}$ as in Ref. \cite{GFCEmergentFriedmann}.

Some remarks are in order. Firstly, the isotropic restriction on the coordinates of the group manifold used here differs somewhat to the one employed in the spin-representation in Ref. \cite{GFCEmergentFriedmann}. Interestingly, the results do only vary slightly from a quantitative point of view suggesting that they describe (almost) the same quantum geometric configuration. Second, in contrast to Ref. \cite{GFClowspin} the condensates used here exclude the occurrence of excitations with $j=0$. In our context these would correspond to the mean field solving the equation of motion for $\mu=0$, which is identical to zero (cf. Ref. \cite{GFClowspinStaticEffInt}). Also, the mean field configurations considered here obey the near-flatness condition, that means the quantum geometric building blocks are almost flat as naively required for the emergence of an effectively smooth and continuous $3$-geometry from them \cite{GFCReview,GFClowspinStaticEffInt}. We also do not restrict ourselves \textit{a priori} to a configuration with just one $j$. Rather we keep the analysis as general as possible and use only in a final step the relaxation mechanism into the $j=1/2$ configuration to recover the Friedmann equations. 

Finally, we want to investigate the question whether in the case of the free model it is possible to obtain an era of accelerated expansion of the space built from the GFT quanta which lasts long enough to replace the standard inflationary scenario by relying entirely on quantum geometric arguments. We do this by following closely the techniques developed in Ref. \cite{GFCEmergentFriedmann2, GFCEmergentFriedmann3} with some slight deviations to account for the characteristics of the model studied here.

A necessary condition to call a possible era of accelerated expansion an inflationary era is that the number of e-folds
\be\label{efolds1}
N=\frac{1}{3}\log\biggl(\frac{V_{\textrm{end}}}{V_{\textrm{beg}}}\biggr)
\ee
must be \footnote{In models with a bounce, as it is the case here, it is in principle sufficient to have a smaller number of e-folds.} $N\gtrsim 60$  where $V_{\textrm{beg}}$ is the volume of the Universe at the beginning of the phase of accelerated expansion and $V_{\textrm{end}}$ denotes its volume at the end of it. To simplify the following calculations, we assume then that the volume receives its major contribution at both points from a single $j$-mode. Motivated by the results of subsection \ref{lowspinisotropic}, we choose $j$ to be equal to $1/2$ and thus set $\sigma_{\frac{1}{2};\frac{1}{2}}(\phi)\equiv\sigma$. Hence Eq. (\ref{efolds1}) is recast into
\be\label{efolds2}
N=\frac{2}{3}\log\biggl(\frac{\sigma_{\textrm{end}}}{\sigma_{\textrm{beg}}}\biggr).
\ee
In a next step, we have to introduce a physically sensible definition of the notion of acceleration given in terms of the relational clock $\phi$. This is motivated via the Raychaudhuri equation for the acceleration in standard cosmology in terms of proper time $t$ which is
\be\label{accelerationequation}
\frac{\ddot{a}}{a}=\frac{1}{3}\biggl[\frac{\ddot{V}}{V}-\frac{2}{3}\biggl(\frac{\dot{V}}{V}\biggr)\biggr].
\ee
Using that the momentum conjugate to the scalar field $\phi$ is given by $\pi_{\phi}=V\dot{\phi}$, together with $\dot{V}=\partial_{\phi}V ~(\pi_{\phi}/V)$, this expression can be rewritten as
\be\label{accelerationexpression}
\frac{\ddot{a}}{a}=\frac{1}{3}\biggl(\frac{\pi_{\phi}}{V}\biggr)^2\biggl[\frac{\partial^2_{\phi}V}{V}-\frac{5}{3}\biggl(\frac{\partial_{\phi}V}{V}\biggr)^2\biggr]\equiv\frac{1}{3}\biggl(\frac{\pi_{\phi}}{V}\biggr)^2\mathfrak{a}(\sigma),
\ee
as given in Ref. \cite{GFCEmergentFriedmann3}.\footnote{We want to remark that using Eq. (\ref{accelerationequation}) from classical cosmology as an input to motivate Eq. (\ref{accelerationexpression}) should be taken with a grain of salt: At the current stage of the GFT condensate program it is not possible to give an intrinsic derivation for the acceleration from within GFT due to the absence of more interesting observables apart from the number operator as well as the area and volume operators imported from LQG.} Using that the expression for the ``GFT energy" Eq. (\ref{GFTEnergyjm}) for such a configuration is
\be
E_{\frac{1}{2};\frac{1}{2}}\equiv E=\frac{1}{2}\biggl[\tau\sigma'^2+(3/2+\mu)\sigma^2\biggr],
\ee
and setting $\sigma'=0$ we yield
\be
\sigma_{\textrm{beg}}^2=\frac{2E}{\mu+\frac{3}{2}}~,
\ee
for which the acceleration condition $\mathfrak{a}(\sigma)>0$ is fulfilled. The acceleration phase comes to an end when the expression for $\mathfrak{a}(\sigma)$ vanishes which gives
\be
\sigma_{\textrm{end}}^2=\frac{7}{2}\frac{E}{\mu+\frac{3}{2}}.
\ee
For these one obtains that the number of e-folds is $N\approx 0.186$. This value agrees with the upper bound found for the complex-valued model considered in Ref. \cite{GFCEmergentFriedmann3}. There is no lower bound on $N$ due to the absence of the conserved charge $Q$ that is only present in complex-valued models. Also the effect of the Laplacian has no influence on the value of $N$. The small value of $N$ shows that the epoch of accelerated expansion for the case of the free model does not last long enough to offer an alternative to the standard inflationary paradigm. However, employing the same techniques we will show in subsection \ref{isotropicacceleratedexpansion} that for two interaction terms with fine-tuned coupling constants an era of inflation driven by quantum geometric effects can in principle be realized.

We want to remark that the analytic result for $N$ is only possible if just one mode for the condensate is considered. Taking into account all modes would require a numerical analysis which is left for future research. However, we do not expect the value of $N$ to grow significantly then since the biggest contribution to $V_{\textrm{end}}$ and $V_{\textrm{beg}}$ will always stem from the fastest growing condensate components.

\subsection{Relational evolution of an isotropic and effectively interacting GFT condensate configuration: Thomas-Fermi approximation}\label{isotropicthomasfermi}

In this subsection we study the dynamics of effectively interacting GFT condensates in the so-called Thomas-Fermi approximation. This is the regime in which the effect of the Laplacian operator in the kinetic term is considered to be much smaller compared to any interaction. In the case of real BECs this is a typical simplification of the system when the density of the ground state is very large \cite{BECs,NumericalrealBECs}. Here, we analyze this regime first by performing a formal stability analysis of the nonlinear dynamical system around its fixed points and second by finding numerically solutions to the equation of motion giving rise to an effective Friedmann equation.

To start with, for a system with one interaction term, the equation of motion
\be
\biggl[(\tau\partial^2_{\phi}-\sum_{I=1}^4\Delta_{g_{I}})+M^2\biggr]\sigma(g_I,\phi)+\kappa \sigma(g_{I},\phi)^{n-1}=0,
\ee
with $n=4,5,...$ under the above-discussed isotropic restriction leads to
\begin{multline}
\biggl[(2\tau\partial^2_{\phi})-(\frac{d^2}{d\psi^2}+2\cot(\psi)\frac{d}{d\psi})+2\mu\biggr]\sigma(\psi,\phi)\\~+~2\kappa \sigma(\psi,\phi)^{n-1}=0.
\end{multline}

When considering the analog of the Thomas-Fermi approximation for real BECs, the contribution of the Laplace-Beltrami operator is suppressed which implies that $\sigma(\psi,\phi)=\xi(\psi) T(\phi)=c~T(\phi)$ with some constant $c$. In Ref.~\cite{GFClowspinStaticEffInt} it was shown that the Fourier components of such constant functions on the domain are dominated by the lowest nontrivial mode which implies that the condensate consists of many smallest possible and discrete building blocks. 

Its dynamics are then captured by the nonlinear dynamical system
\be
(\tau\partial^2_{\phi}+\mu)T(\phi)+\kappa T(\phi)^{n-1}=0
\ee
where factors of $c$ were absorbed into $\kappa$. With $T(\phi)\equiv x$ and $T'(\phi)\equiv y$ and $\tau\equiv 1$ this can be rewritten as
\begin{equation}\label{dynamicalsystem1}
\frac{d}{d\phi} 
 \begin{pmatrix}
  x \\
  y
 \end{pmatrix}
 =
 \begin{pmatrix}
  y \\
  -\mu x-\kappa x^{n-1}
 \end{pmatrix}.
\end{equation}
For even-valued $n$, the formal linear stability analysis of its fixed points $(x,y)_{*}=\{(0,0),(\pm\sqrt[n-2]{-\mu/\kappa},0)\}$ shows that for $\mu<0$ and $\kappa>0$ the first is a saddle and, the others are center fixed points. For odd-valued $n$, only for $\mu<0,~\kappa>0$ or $\mu<0,~\kappa<0$ one can have a nontrivial (nonperturbative) vacuum in agreement with the condensate state ansatz, so the fixed points of the dynamical system are given by $(x,y)_{*}=\{(0,0),(\pm\sqrt[n-2]{\mp\mu/\kappa},0)\}$. 

Using this, we can give the general solutions close to the fixed points. For example, in the case for an even-valued $n$, close to the saddle point, the solution is given by
\begin{equation}\label{linearizedsolutionsaddle}
 \begin{pmatrix}
  T \\
  T'
 \end{pmatrix}(\phi)
 =
 \frac{\textrm{a}_1 e^{-\sqrt{|\mu|}\phi}}{\sqrt{1+\frac{1}{|\mu|}}}
 \begin{pmatrix}
  -\frac{1}{\sqrt{|\mu|}} \\
  1
 \end{pmatrix}
 +\frac{\textrm{a}_2 e^{\sqrt{|\mu|}\phi}}{\sqrt{1+\frac{1}{|\mu|}}}
 \begin{pmatrix}
  \frac{1}{\sqrt{|\mu|}} \\
  1
 \end{pmatrix},
\end{equation}
where $\textrm{a}_1$ and $\textrm{a}_2$ are determined by using the initial conditions. Around the two center fixed points, the linearized solution is given by
\begin{multline}
 \begin{pmatrix}
  T \\
  T'
 \end{pmatrix}(\phi)
 =
 \begin{pmatrix}
 \pm\sqrt[n-2]{|\mu|/\kappa} \\
  0
 \end{pmatrix}
 \\
+~ \frac{\textrm{a}_1 e^{-i\sqrt{(n-2)|\mu|}\phi}}{\sqrt{1+\frac{1}{((n-2)|\mu|)^2}}}
 \begin{pmatrix}
  \frac{i}{\sqrt{(n-2)|\mu|}}\\
  1
 \end{pmatrix}
 \\+~ \frac{\textrm{a}_2 e^{i\sqrt{(n-2)|\mu|}\phi}}{\sqrt{1+\frac{1}{((n-2)|\mu|)^2}}}
\begin{pmatrix}
  -\frac{i}{\sqrt{(n-2)|\mu|}}\\
  1
 \end{pmatrix}.
\end{multline}

\begin{figure}[ht]
	\centering
  \includegraphics[width=0.4\textwidth]{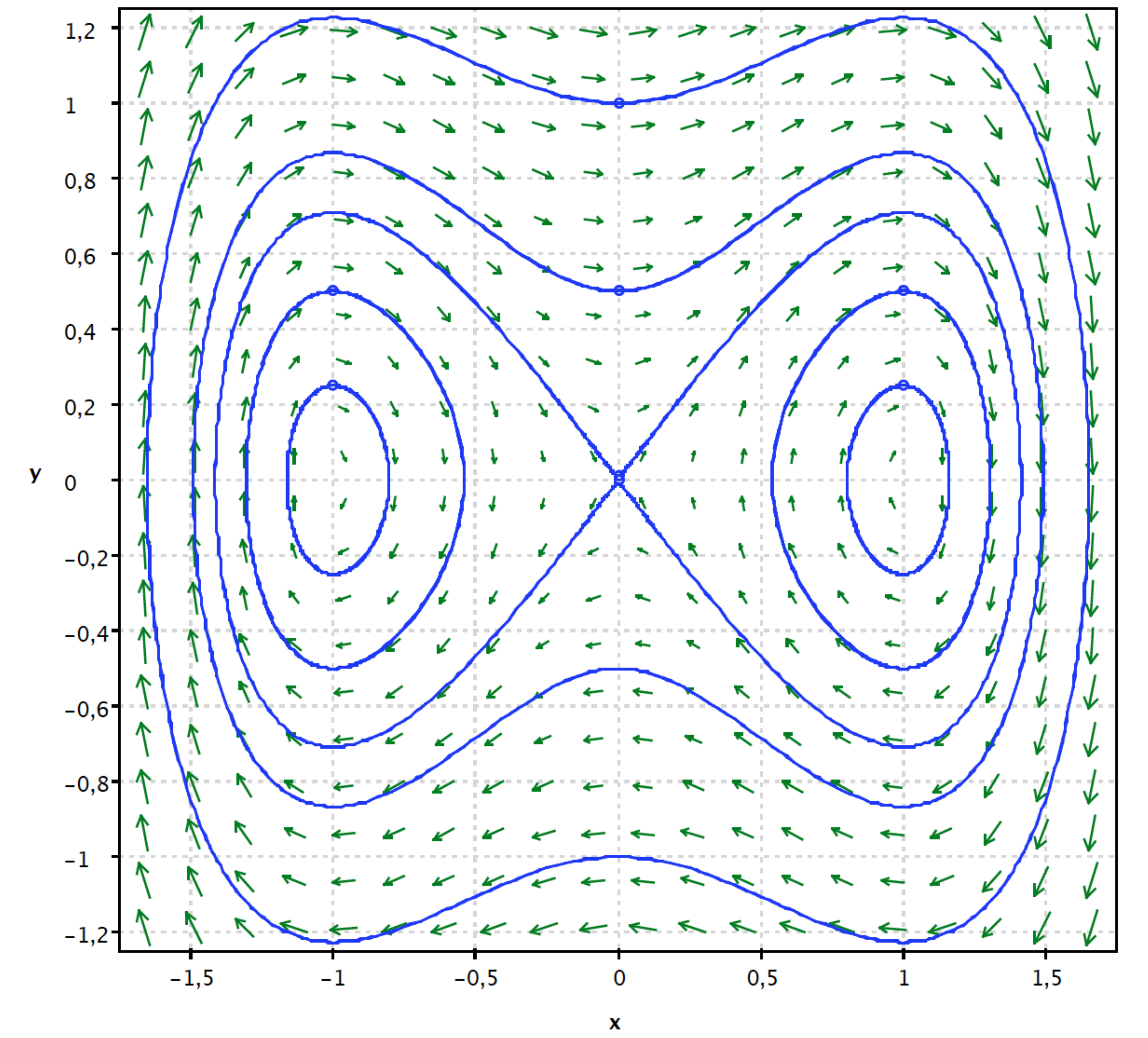}
	\caption{Phase plane portrait of the dynamical system given by Eq. (\ref{dynamicalsystem1}) for $\mu=-1$, $\kappa=1$, $n=4$ and for different initial conditions where $x=T(\phi)$ and $y=T'(\phi)$.}
	\label{figphaseportrait1}
\end{figure}

\begin{figure}[ht]
	\centering
  \includegraphics[width=0.4\textwidth]{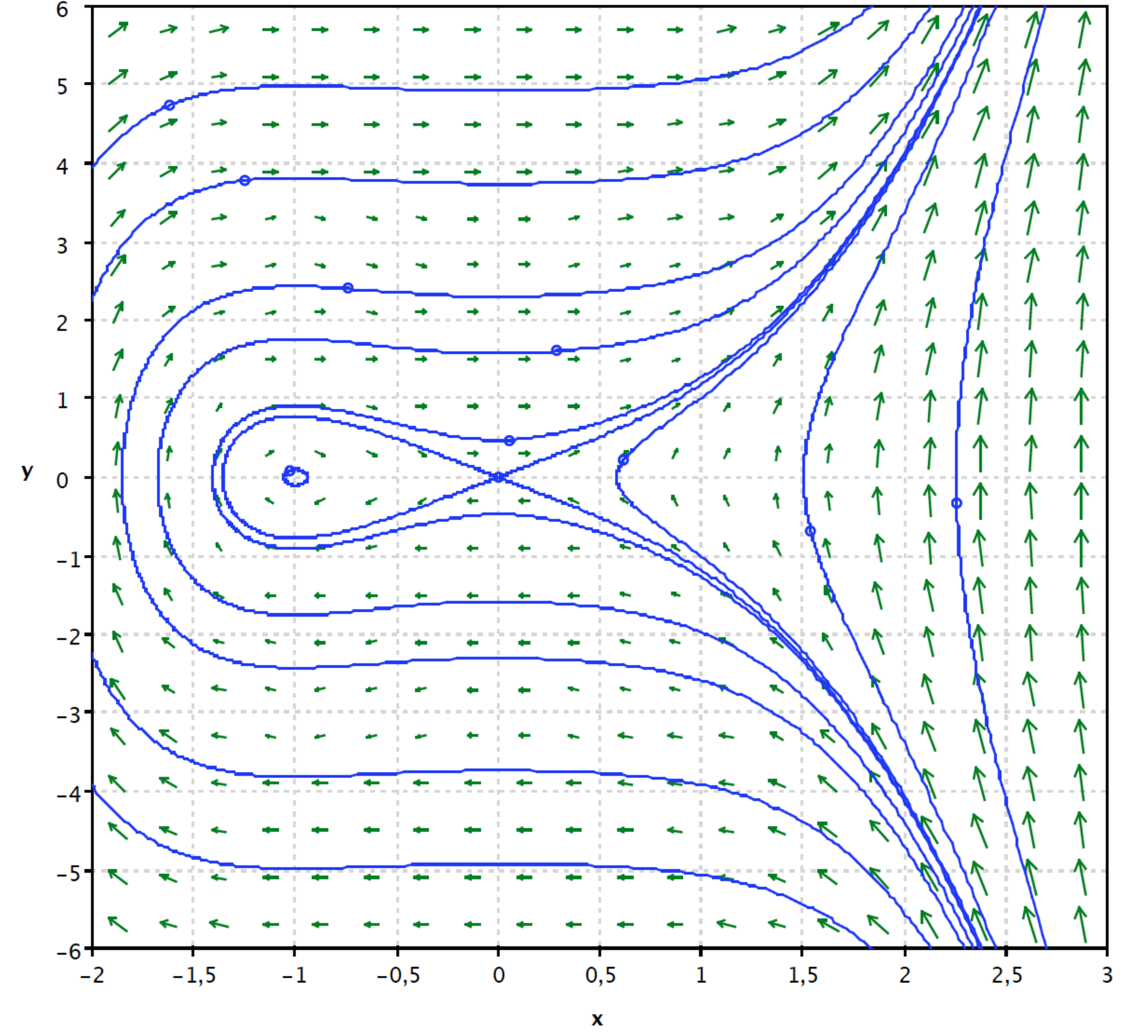}
	\caption{Phase plane portrait of the dynamical system given by Eq. (\ref{dynamicalsystem1}) for $\mu=-1$, $\kappa=-1$, $n=5$ and for different initial conditions where $x=T(\phi)$ and $y=T'(\phi)$.}
	\label{figphaseportrait2}
\end{figure}

The full numerical solutions obtained by employing an implementation of the Runge-Kutta method at small step size are illustrated in the phase portraits given by Figs.~\ref{figphaseportrait1} and \ref{figphaseportrait2} which resemble those of a classical point particle in the corresponding effective potential. From the phase portraits we also deduce that depending on the initial conditions one can find configurations whose total volume vanishes at some point and others for which the volume never becomes zero.

When comparing Eq. (\ref{linearizedsolutionsaddle}) to Eq. (\ref{zeitanteil}) we also notice that close to the saddle point, i.e. where the nonlinearities are neglected, the solutions show a similar evolution behavior with respect to the relational clock. It is thus obvious that only in the neighborhood of this point the effective Friedmann equation
\be
\Biggl(\frac{V'}{3 V}\biggr)^2=\frac{4\pi G}{3},
\ee
can be recovered when using Eq. (\ref{linearizedsolutionsaddle}) together with the identification $3\pi G=\frac{|\mu|}{\tau}$. In the latter identification of course no contribution stemming from the Laplacian is taken into account since we work in the Thomas-Fermi regime. Another difference worth to be emphasized lies then in the fact that to recover the effective Friedmann equations in subsection \ref{emergentfriedmanndynamics} the quick relaxation of the system into a low-spin phase was needed whereas here the system is dominated by the $1/2$-configuration from the onset. 

Away from the fixed point, however, the effective interactions will modify the evolution process due to the nonlinear interactions which is clearly visible from the phase portraits. For instance, for even-powered potentials the evolution will be cyclic in general whereas for odd-powered potentials cyclic evolution competes with an open evolution depending on the initial conditions as Fig.~\ref{figphaseportrait2} illustrates.

More specifically, we can give the dynamical equation in the effectively interacting case for the volume which takes the form of a modified Friedmann equation in relational terms. For this, we can exploit that $\sigma$ is constant on the domain in the Thomas-Fermi regime, which simplifies the expression for the ``GFT energy" in Eq. (\ref{GFTEnergy}) to
\be
E=\frac{1}{2}\biggl[\sigma'^2+\mu\sigma^2+2\frac{\kappa}{n}\sigma^{n}\biggr],
\ee
where we included the interaction term. Since the constancy of the mean field on the domain implies that its dominant Fourier component is $\sigma_{\frac{1}{2};\frac{1}{2}}$, the volume can be expressed as $V=V_{\frac{1}{2}}\sigma_{\frac{1}{2};\frac{1}{2}}(\phi)^2$. Thus the first Friedmann equation in terms of the relational clock is given by
\begin{multline}
\biggl(\frac{V'}{3 V}\biggr)^2=
\frac{4\pi G}{3}+\frac{4}{9}\biggl[\frac{2E V_{\frac{1}{2}}}{V}-\frac{2\kappa}{n}\biggl(\frac{V_{\frac{1}{2}}}{V}\biggr)^{1-n/2}\biggr],
\end{multline}
where we used the identification $|\mu|=3\pi G$. As stated above, dimensional analysis suggests that the first term in the parenthesis is of the order of $\sqrt{\hbar G}$ and is thus a quantum gravity correction (cf. Ref. \cite{GFCEmergentFriedmann}). Since the units of GFT coupling constants are not clear from a fundamental point of view and have to be consistent with (semi-)classical results, similarly, since $V_{\frac{1}{2}}\sim (\hbar G)^{3/2}$ then only if $\kappa$ is of the order of $(\hbar G)^x$ with $x>\frac{3}{4}n-\frac{3}{2}$, the second term stemming from the GFT interaction term can be understood as a quantum gravity correction which vanishes in the limit $\ell_p\to 0$.


\subsection{Relational evolution of an isotropic and effectively interacting GFT condensate configuration: Geometric inflation}\label{isotropicacceleratedexpansion}

In the following we demonstrate that within the context of the effectively interacting GFT condensate models it is in principle possible to have an era of accelerated expansion lasting long enough to represent an alternative to the standard inflationary scenario. This, however, comes at the cost of using two interaction terms and a fine-tuning of the respective coupling constants. The origin of these results is entirely quantum geometric since the massless scalar field $\phi$ used throughout this article serves only as a relational clock and does not play the role of the inflaton.

To this aim, we will employ the techniques to extract information about the acceleration encoded by a model with a complex-valued GFT field in the spin-representation that were developed in Ref. \cite{GFCEmergentFriedmann3}. Despite the fact that here we work with a real-valued model starting in the group-representation and in principle different effective interactions, within the Thomas-Fermi regime we are able to fully recover the same results as in Ref. \cite{GFCEmergentFriedmann3}. This is also the reason why we keep the subsequent exposition as brief as possible.

As argued in subsection \ref{isotropicthomasfermi}, in the Thomas-Fermi regime the expectation value of the volume operator can be written as $V=V_{\frac{1}{2}}\sigma_{\frac{1}{2};\frac{1}{2}}(\phi)$ and we drop indices of the mean field hereafter. Including the ``mass term" and two interaction terms, the effective potential reads
\be
V_{\textrm{eff}}=\frac{\mu}{2}\sigma^2+\frac{\kappa}{n}\sigma^n+\frac{\lambda}{n'}\sigma^{n'}
\ee
wherein for stability purposes we choose $\lambda>0$ and $n'$ to be even. For the ``GFT energy" we can then write
\be
E=\frac{1}{2}[\sigma'^2+2 V_{\textrm{eff}}].
\ee
With this we can cast the expression for the ``acceleration" $\mathfrak{a}(\sigma)$ given by Eq. (\ref{accelerationexpression}) first into
\be
\mathfrak{a}(\sigma)=-\frac{2}{\sigma^2}\biggl[\frac{(\partial_{\phi}V_{\textrm{eff}})\sigma}{\sigma'}+\frac{14}{3}(E-V_{\textrm{eff}})\biggr]
\ee
which is then rewritten as
\be\label{accelerationinteracting}
\mathfrak{a}(\sigma)=-\frac{2}{3\sigma^4}\biggl[-4\mu\sigma^4+14 E\sigma^2+\alpha\sigma^{n+2}+\beta\sigma^{n'+2}\biggr]
\ee
with $\alpha=(3-14/n)\kappa$ and $\beta=(3-14/n')\lambda$.

As stated above, in order to accommodate an era of inflationary expansion we assume the hierarchy $|\kappa|\gg\lambda$ hereafter which amounts to a fine-tuning. The beginning of the acceleration phase starts when $E=V_{\textrm{eff}}$ where the acceleration condition $\mathfrak{a}(\sigma_{\textrm{beg}})>0$ must hold. This leads to the condition $\partial_{\phi}V_{\textrm{eff}}\leq 0$, which together with the hierarchy implies $|\mu|\geq \kappa$.  Furthermore, we have to demand that $\alpha<0$ so that a large enough number of e-folds can be generated in this model, since otherwise $\sigma_{\textrm{end}}$ is even smaller than in the free case treated in subsection \ref{emergentfriedmanndynamics}.

At the end of the accelerated expansion phase the acceleration has to vanish, i.e. $\mathfrak{a}(\sigma)=0$, from which we obtain $\sigma_{\textrm{end}}$. To do so we can expect that $\sigma_{\textrm{end}}\gg \sigma_{\textrm{beg}}$, which allows us to solve for the root of $\mathfrak{a}$ by considering solely the highest powers in $\sigma$ therein. With this one can argue that in order to guarantee that $\sigma_{\textrm{end}}$ is the only root of $\mathfrak{a}$ in between $\sigma_{\textrm{beg}}$ and $\sigma_{\textrm{end}}$ one needs $\kappa<0$. Together with $\sigma_{\textrm{end}}=\sigma_{\textrm{beg}}~e^{\frac{3}{2}N}$ and the expressions for $\alpha$ and $\beta$ this yields
\be
\beta=-\alpha~e^{-\frac{3}{2}N(n'-n)}>0.
\ee
From this, the value of $N$ can be obtained when the values of $\kappa$, $n$ and $\lambda$, $n'$ are given. It also implies that $n'>n\geq 5$. 

This is interesting because for $n=5$ the corresponding local interaction term would mimic a proper combinatorially nonlocal interaction term which is typically needed to formulate a GFT with a simplicial quantum gravity interpretation. Since $n'$ has to be even for stability reasons, the corresponding even-powered interaction terms are reminiscent of so-called tensorial interactions. Typically, these occur in models where the GFT field is endowed with a specific tensorial transformation property (cf. \cite{GFTReview}). We refrain from giving further phenomenological arguments given in Ref. \cite{GFCEmergentFriedmann3} which indicate that only $n'=6$ is physically reasonable because those arguments applied to the model considered there would go through in just the same way here.

We want to emphasize that the above given strategy to extract the acceleration behavior of the emergent space as described by the mean field $\sigma$ is within our context only applicable to the case of the Thomas-Fermi regime in which the mean field is constant on the domain. To go beyond this and study less restricted scenarios, one would have to employ numerical techniques which can control the regular singularities of the Laplacian arising due to the used coordinate system. This is left to future research. However, it is interesting that the same acceleration pattern is generated by means of real- and complex-valued models where the powers of the effective interaction terms agree whereas the microscopic details of their effect on individual GFT quanta is rather different. 

\subsection{Relational evolution of an isotropic and effectively interacting GFT condensate configuration: Formal stability properties of the full system}\label{isotropiclinearizedsystem}

In a final step we want to study the stability of the condensate system close to the fixed points computed in subsection \ref{isotropicthomasfermi} when the effect of the Laplace-Beltrami operator onto the system is also included. To this aim, we employ a linearization of the nonlinear PDE about these points which is the standard procedure used in the context of the stability theory of general dynamical systems. In a first step, without loss of generality, we set $\tau\equiv 1$ and cast the original system into first order form
\be
\frac{d}{d\phi} 
 \vec{\Sigma}
 \equiv
\frac{d}{d\phi} 
 \begin{pmatrix}
  \sigma \\
  \sigma'
 \end{pmatrix}
 =
 \begin{pmatrix}
  \sigma' \\
  \Delta\sigma-\mu \sigma-\kappa \sigma^{n-1}
 \end{pmatrix}
\ee
with stationary solutions
\be
\vec{\Sigma}_{*}(\psi)=
\begin{pmatrix}
  \sigma(\psi) \\
  0
 \end{pmatrix}.
\ee
Plugging the ansatz $\vec{\Sigma}=\vec{\Sigma}_{*}+\vec{\Omega}$ into the original nonlinear equation, where $\vec{\Omega}$ represents a small perturbation about the fixed points, one finds the linearization of the nonlinear PDE at the solution $\vec{\Sigma}_{*}$ to be given by
\be
\frac{d}{d\phi}\vec{\Omega}=J_{\vec{\Sigma}_{*}}\vec{\Omega}+\mathcal{O}(\vec{\Omega}^2),
\ee
where $J$ denotes the Jacobian. This gives
\be
\frac{d}{d\phi}\vec{\Omega}=
\begin{pmatrix}
  0 & 1\\
  \Delta-\mu-\kappa(n-1)\sigma(\psi)^{n-2} & 0
 \end{pmatrix}\vec{\Omega}.
\ee
For any initial condition its solution is given by
\be
\vec{\Omega}=e^{J_{\vec{\Sigma}_{*}}\phi}\vec{\Omega}_{0}.
\ee
The solution is called stable if for its eigenvalues $\lambda_i$ with $i\in\{1,2\}$ of $J_{\vec{\Sigma}_{*}}$ one has $\operatorname{Re}(\lambda_i)\leq 0$. Otherwise it is called unstable. 

For the saddle point $\vec{\Sigma}_{*}=(0,0)$ the eigenvalues of the Jacobian read
\be
\lambda_{1,2}=\pm\sqrt{-2j(j+1)+|\mu|}.
\ee
Stable solutions are only found if and only if these eigenvalues are purely imaginary. Due to the linearization, we can see when this happens by directly importing the results of subsection \ref{lowspinisotropic}. This allows us to reinterpret the blue sector in Fig.~\ref{figparameterlowspin} as only comprising of stable solutions whereas the orange sector represents unstable ones. The striking fact that small $j$-modes are more unstable than all the others, leading to their exponential growth, has a simple explanation from the point of view of the formal stability analysis. For a given $\mu$, only for large enough $j$ the contributions stemming from the Laplacian and the negative ``mass term" will be positive altogether. Hence, for such modes, the equation of motion resembles that of a particle in a potential which is bounded from below. When $j$ is too small to compensate for the negative ``mass term", the latter would appear as unbounded, wherefore these modes are unstable. These are the physically most relevant ones, as they lead to a quick expansion of the emergent space.

In contrast, for the center fixed points $\vec{\Sigma}_{*}=(\pm\sqrt[n-2]{-\mu/\kappa},0)$ the eigenvalues of the Jacobian are
\be
\lambda_{1,2}=\pm\sqrt{-2j(j+1)+(2-n)|\mu|}
\ee
which for $n=4,6,...$ always leads to stable solutions. 

Supposing that one chooses the vicinity of the saddle point as the starting point of the evolution of the condensate, the latter will roll down the effective potential towards the local minimum and will dynamically settle into a low-spin configuration. Depending on the value of $\mu$ this leads to a condensate configuration where each building block of the quantum geometry is only characterized by the mode for which $j$ is equal to $1/2$.\footnote{Similar results hold for odd-powered potentials, only that these exhibit just one center fixed point apart from the saddle point.}

\section{Relational evolution of an anisotropic GFT condensate configuration}\label{anisotropies}

In the subsequent section, we will lift the isotropic restriction which we recapitulated in subsection \ref{recapstaticfree}, to pave the way to study for the first time more general, i.e. anisotropic GFT condensate configurations from the point of view of the group-representation. The purpose of the following exposition is to serve as a starting point for a more systematic analysis of anisotropic GFT condensate configurations, their dynamics, comparison to the study of anisotropic models in LQC \cite{LQC} as well as spin foam cosmology and the possible emergence of generalized Friedmann equations (cf. \cite{CosmologicalModels}) in an appropriate limit. 

More precisely, after explaining how such configurations can be obtained, we study their behavior in the small and large volume regimes. We show that anisotropies are dominant in the former and play a negligible role in the latter regime. We conclude this section by performing a formal stability analysis of the corresponding effectively interacting GFT condensate system which explores and explains the reasons for such a behavior. To this aim, we use again techniques used in the context of the stability theory of general dynamical systems.

\subsection{Relational evolution of an anisotropic and free GFT condensate configuration: Dynamical isotropization}\label{anisoiso}

At the beginning of Section \ref{sectionstatic}, we introduced coordinates on the group $\mathrm{SU(2)}$ and expressed the Laplace-Beltrami operator in terms of these before employing an isotropic restriction onto the condensate wave function in subsection \ref{recapstaticfree}. In the present subsection, however, we will refrain from the full symmetry reduction and retain some of the anisotropic information which is stored in the mean field. Using such a particular anisotropic configuration, we will show in a similar manner as in the previous subsection \ref{lowspinisotropic} that by dynamically settling into a low-spin phase, the condensate isotropizes over time.

To this aim, we come back to the expression of the full Laplace-Beltrami operator, i.e. Eq. (\ref{kinetictermLaplacian}), and consider a particular geometry of the quantum tetrahedron. In general, the fluxes $\vec{B}_i$ associated to the tetrahedron are perpendicular to its faces. Bearing this in mind, let us assume that we are working with a trirectangular tetrahedron which has three pairwisely orthogonal faces. Hence, for such a building block of the quantum geometry
\be\label{trirectangulartetrahedron}
\sum_{i\neq j}\vec{B}_i\cdot\vec{B}_j=0
\ee
holds. Imposing such a constraint, for a mean field which depends only on the diagonal components $\pi_{II}$ (thus $p_1, p_2, p_3$) and the relational clock $\phi$, this frees us from the second sum in Eq. (\ref{laplacefirstlevelreduction}) and allows us to decouple all remaining coordinates from one another.

It should be noticed that Eq. (\ref{trirectangulartetrahedron}) implies the vanishing of the off-diagonal components in the reconstructed metric, Eq. (\ref{reconstructedmetric}). Hence, this restriction suggests that the anisotropic condensate states to be constructed now can possibly be related to the Bianchi models of class A (cf. \cite{GFCReview, CosmologicalModels}).

To be concrete, using Eq. (\ref{kinetictermLaplacian}) and (\ref{trirectangulartetrahedron}) the resulting equation of motion leads to the following PDE for $\sigma(p_1, p_2, p_3, \phi)$
\begin{multline}\label{PDEdecoupled}
\biggl[\tau\partial_{\phi}^2-\sum_{i=1}^3(2p_i(1-p_i)\frac{d^2}{dp_i^2}~+~\\(3-4p_i)\frac{d}{dp_i})+\mu\biggr]\sigma(p_1,p_2,p_3,\phi)=0.
\end{multline}
Since our assumptions have decoupled all the terms in the $p_i$'s and $\phi$ from each other, we can employ a separation ansatz to solve Eq. (\ref{PDEdecoupled}). Together with the substiution $p_i=\sin^2(\psi_i)$ the product ansatz
\be
\sigma(\psi_1,\psi_2,\psi_3,\phi)=\xi(\psi_1)\xi(\psi_2)\xi(\psi_3)T(\phi)
\ee
yields
\begin{eqnarray}
2\tau\frac{\partial_{\phi}^2T(\phi)}{T(\phi)}&=&\sum_{i}\frac{(\frac{d^2}{d\psi_i^2}+2\cot(\psi_i)\frac{d}{d\psi_i})\xi(\psi_i)}{\xi(\psi_i)}-2\mu~\nonumber\\
&\equiv&~\omega=\mathrm{const}.
\end{eqnarray}
The general solution to the equation for $T(\phi)$ is again given by
\be\label{zeitentwicklung2}
T(\phi)=\biggl(\mathrm{a}_1~e^{\sqrt{\frac{\omega}{2\tau}}\phi}+\mathrm{a}_2~e^{-\sqrt{\frac{\omega}{2\tau}}\phi}\biggr),
\ee
with the constants $\mathrm{a}_1,\mathrm{a}_2\in\mathbb{C}$.

Using $\mu=\sum_{i=1}^3\mu_i$ and $\omega=\sum_{i=1}^3\omega_i$ one solves in each direction the one-dimensional problem
\be
\biggl[-(\frac{d^2}{d\psi_i^2}+2\cot(\psi_i)\frac{d}{d\psi_i})+2\mu_i+\omega_i\biggr]\xi(\psi_i)=0,~~\psi_i\in[0,\frac{\pi}{2}],
\ee
separately. To solve these differential equations, we impose, as above, Dirichlet boundary conditions $\xi(\psi_i=\frac{\pi}{2})=0$. Then the eigensolutions are given by
\be
\xi_{j_{i}}(\psi_i)=\frac{\sin((2j_i+1)\psi_i)}{\sin(\psi_i)},~~\psi_i\in[0,\frac{\pi}{2}]
\ee
with eigenvalues $\omega_i+2\mu_i=-4j_i(j_i+1)$ and $j\in\frac{2\mathbb{N}_0 +1}{2}$.

Using the above, we get
\be
\omega=-2\mu-4\sum_i{j_i(j_i+1)}
\ee
which  substituting into Eq. (\ref{zeitentwicklung2}) yields
\begin{multline}
T_{j_1,j_2,j_3}(\phi)=\\~\biggl(\mathrm{a}_1~e^{\sqrt{\frac{\mu+2\sum_i j_i(j_i+1)}{-\tau}}\phi}+\mathrm{a}_2~e^{-\sqrt{\frac{\mu+2\sum_i j_i(j_i+1)}{-\tau}}\phi}\biggr).
\end{multline}
As in the previous section, its behavior critically depends on the sign of $\frac{\mu+2\sum_i j_i(j_i+1)}{-\tau}$ where $\mu<0$ and $\tau>0$. Assuming now that $\mu_i<0$, solutions for which 
\be
\frac{1}{2}\leq j_i < -\frac{1}{2}+\frac{1}{2}\sqrt{1-2\mu_i}~~\mathrm{with}~~\mu_i< -\frac{3}{2} 
\ee
grow exponentially as $\phi\to\pm\infty$ for $i=1,2,3$ simultaneously. Strikingly, the condensate will quickly be dominated by the lowest representation $j_i=\frac{1}{2}$ for all $i=1,2,3$ and all other $j$-contributions are suppressed. This directly implies that toward $\phi\to\pm\infty$ the particular anisotropic configuration
\begin{multline}
\sigma(\psi_1,\psi_2,\psi_3,\phi)~=~\\\sum_{j_1, j_2, j_3\in\frac{2\mathbb{N}_0+1}{2}}\xi_{j_1}(\psi_1)\xi_{j_2}(\psi_2)\xi_{j_3}(\psi_3)T_{j_1,j_2,j_3}(\phi)
\end{multline}
will dynamically isotropize, as indicated by the late time behavior of the expectation value of the volume operator
\begin{multline}
\lim_{\phi\to\pm\infty}V(\phi)~=~\\V_0~V_{1/2}~\biggl(|\xi_{1_{\frac{1}{2};\frac{1}{2}}}|^2\biggr)^3~|\mathrm{a}_{1,2}|^2~e^{\pm 2\sqrt{\frac{\mu+\frac{9}{2}}{-\tau}}\phi}.
\end{multline}
Consequently, a generalized Friedmann equation (in relational terms) which would take into account the effect of the anisotropies, will reduce at late times to the one for the isotropic configuration.

\begin{figure}[ht]
	\centering
  \includegraphics[width=0.4\textwidth]{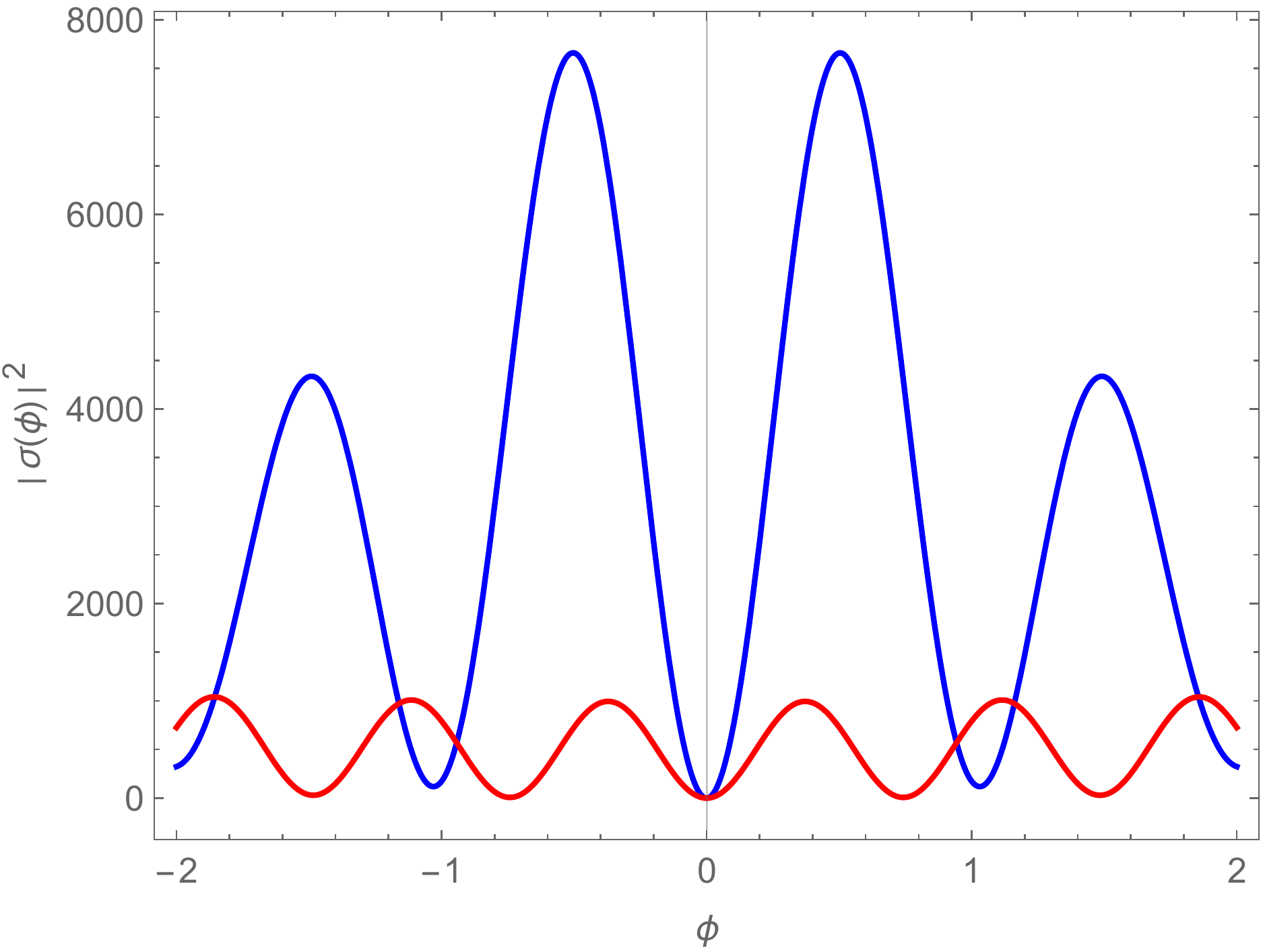}
	\caption{Probability density $|\sigma(\phi)|^2$ of the mean field for the isotropic (red) and the anisotropic (blue) parts for small values of the relational clock $|\phi|$ for $\mu=-4.6$ and $j_{\mathrm{max}}=3/2$. Setting $j_{\mathrm{max}}$ to higher values does not qualitatively change the result.}
	\label{figisovsanisofirst}
\end{figure}

\begin{figure}[ht]
	\centering
  \includegraphics[width=0.4\textwidth]{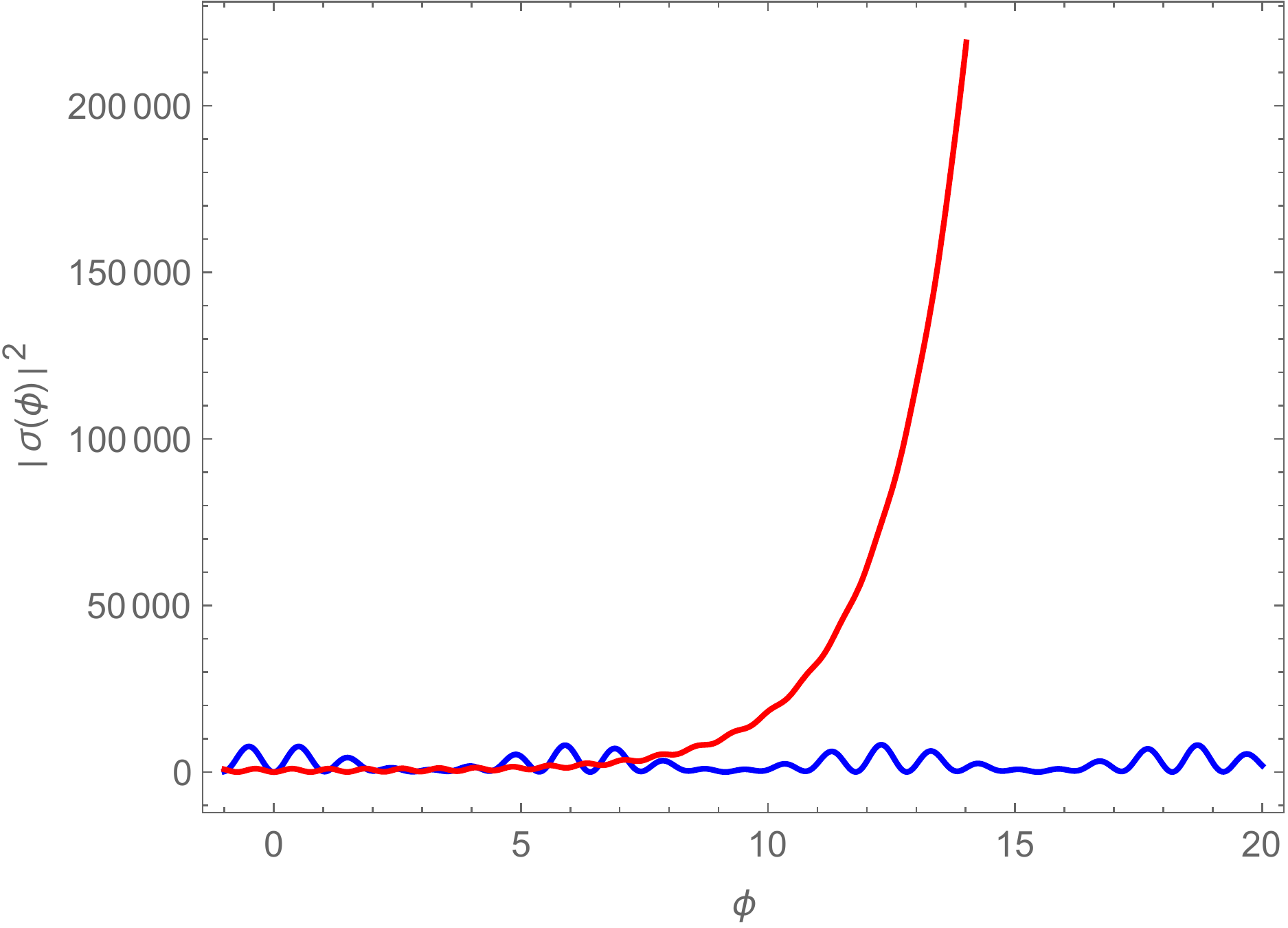}
	\caption{Probability density $|\sigma(\phi)|^2$ of the mean field for the isotropic (red) and the anisotropic (blue) parts for larger values of the relational clock $\phi$ for $\mu=-4.6$ and $j_{\mathrm{max}}=3/2$. Setting $j_{\mathrm{max}}$ to higher values does not qualitatively change the result.}
	\label{figisovsanisosecond}
\end{figure}

In a last step, we want to investigate whether the contributions stemming from the anisotropies become important at small volumes. To this aim, let us consider without loss of generality the initial condition $T_{j_1,j_2,j_3}(0)=0$.\footnote{More generally, one could also use that Eq. (\ref{zeitentwicklung2}) vanishes for $\phi=\sqrt{\frac{2\tau}{\omega}}\frac{\log(-\mathrm{a}_1/\mathrm{a}_2)}{2}$. However, this would not change the subsequent discussion qualitatively.} For this one has
\begin{multline}
T_{j_1,j_2,j_3}(\phi)=2\mathrm{a}_1\sinh\biggl(\sqrt{\frac{\mu+2\sum_i j_i(j_i+1)}{-\tau}}\phi\biggr).
\end{multline}
The differences between the dynamical behavior of the isotropic ($j_1=j_2=j_3$) and the anisotropic part of the mean field are illustrated by Figs. \ref{figisovsanisofirst} and \ref{figisovsanisosecond}. These show that anisotropies only play an important role at small values of the relational clock, i.e. small volumes, whereas at late times the isotropic mode for $j_1=j_2=j_3=1/2$ will clearly dominate. Certainly, this behavior is qualitatively the same for the other branch of solutions (with initial conditions $T_{j_1,j_2,j_3}'(0)=0$) where the singularity problem is avoided since $\lim_{\phi\to 0}T_{j_1,j_2,j_3}(\phi)\neq 0$. We also want to remark that in spite of the surge of anisotropies for small volumes, towards $\phi\to 0$ such a behavior cannot turn a solution corresponding to a finite volume into one for which the volume vanishes, and vice versa.


From a physical point of view, such a behavior is certainly very interesting calling for an explanation. In the ensuing subsection, this will be given by means of a formal stability analysis of the corresponding dynamical system around its fixed points which shows that the solutions for isotropic modes with $j_1=j_2=j_3=1/2$ are more unstable as compared to others. These are the physically most relevant modes, since they lead to a quick and isotropic expansion of the emergent geometry.

Given that our discussion rests on a particular anisotropic building block, it would be interesting to lift this restriction, considering even more general configurations and investigating whether such a behavior is then also realized. Moreover, it would be important to systematically relate and compare such investigations to anisotropic quantum cosmological models explored in LQC \cite{LQC} which will be done elsewhere.

\subsection{Relational evolution of an anisotropic and effectively interacting GFT condensate configuration: Formal stability properties of the full system}\label{anisostability}

We want to analyze the formal stability properties of the anisotropic and effectively interacting condensate system in the vicinity of the fixed points of its dynamics. This is very analogous to the corresponding analysis for the isotropic configuration which is why we keep the subsequent argumentation as short as possible.  

Again, we employ a linearization of the nonlinear PDE about the fixed points, however, keeping in mind that the mean field depends now on three coordinates on the domain, as well as on the relational clock, i.e. $\sigma=\sigma(\psi_1,\psi_2,\psi_3,\phi)$. Without loss of generality, we set $\tau\equiv 1$ and rewrite the original equation of motion into first order form
\be
\frac{d}{d\phi} 
 \vec{\Sigma}
 \equiv
\frac{d}{d\phi} 
 \begin{pmatrix}
  \sigma \\
  \sigma'
 \end{pmatrix}
 =
 \begin{pmatrix}
  \sigma' \\
  \Delta\sigma-\mu \sigma-\kappa \sigma^{n-1}
 \end{pmatrix}
\ee
which has the stationary solutions
\be
\vec{\Sigma}_{*}(\psi_1,\psi_2,\psi_3)=
\begin{pmatrix}
  \sigma(\psi_1,\psi_2,\psi_3) \\
  0
 \end{pmatrix}.
\ee

With the ansatz $\vec{\Sigma}=\vec{\Sigma}_{*}+\vec{\Omega}$ one finds the linearization of the nonlinear PDE for the anisotropic system at the stationary solutions $\vec{\Sigma}_{*}$ to be given by
\be
\frac{d}{d\phi}\vec{\Omega}=J_{\vec{\Sigma}_{*}}\vec{\Omega}+\mathcal{O}(\vec{\Omega}^2),
\ee
with the Jacobian $J$.

\begin{figure}[ht]
	\centering
  \includegraphics[width=0.35\textwidth]{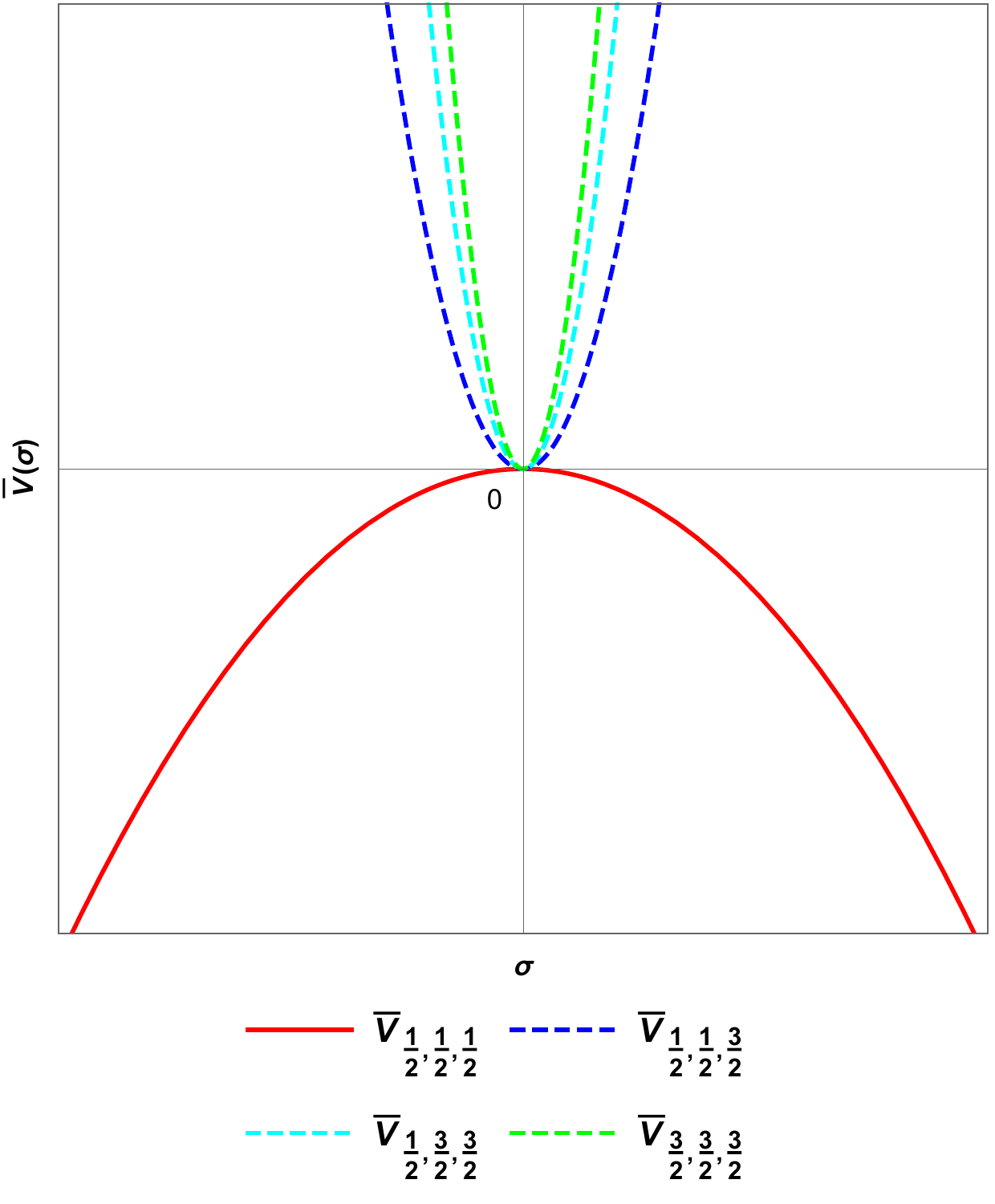}
	\caption{Form of the potentials $\bar{V}_{j_1,j_2,j_3}[\sigma]$ for $\mu=-5$.}
	\label{figpotentials}
\end{figure}

The eigenvalues of the Jacobian at the saddle point $\vec{\Sigma}_{*}=(0,0)$ are given by
\be
\lambda_{1,2}=\pm\sqrt{-2\sum_i j_i(j_i+1)+|\mu|}.
\ee
Only if the eigenvalues are purely imaginary, the corresponding solutions are stable. From this expression we see that this is the case if $j_1,j_2$ and $j_3$ exceed a certain value for a given $\mu$. For such modes the equation of motion resembles that of a particle in a potential
\be
\bar{V}_{j_1,j_2,j_3}[\sigma]=\biggl[2\sum_i j_i(j_i+1)+\mu\biggr]\sigma^2
\ee
which is then bounded from below. When the $j_i$ are too small to compensate for the negative ``mass term", the latter would appear as unbounded, wherefore these modes are considered as unstable from the point of view of the formal stability analysis. This leads to the exponential growth of the corresponding modes. In Fig. \ref{figpotentials} we illustrate the form of $\bar{V}$ for a particular value for $\mu$.

To conclude, using the terminology of dynamical systems, the specific configuration with $j_1=j_2=j_3=1/2$ will be the most unstable one compared to all the others for a given $\mu$, which implies that an initially anisotropic GFT condensate will quickly isotropize by settling into the low-spin phase. On the other hand, anisotropies are more exposed for this reason at small volumes.

\section{Discussion and conclusion}\label{Discussion}

The goal of this article was to investigate the relational evolution of effectively interacting GFT quantum gravity condensate systems describing the dynamics of effective $3$-geometries and to study the tentative quantum cosmological consequences.

To this aim, we worked with real-valued GFT fields in the group-representation and thus adopted a different point of view compared to Refs. \cite{GFCEmergentFriedmann,GFCEmergentFriedmann2,GFClowspin,GFCEmergentFriedmann3} which rest on working in the spin-representation with complex fields. Based on this, in a first step we elaborated further on the static case of a free condensate configuration in an isotropic restriction \cite{GFClowspinStaticEffInt} and computed its noncommutative Fourier transform. This is crucial since it in principle allows to reconstruct the metric from the information encoded by the condensate mean field.

We then moved on to study the evolution of such a configuration with respect to a relational clock. In close analogy to and expanding on the results presented in Ref. \cite{GFClowspin}, we showed for our model that it quickly settles into a lowest-spin configuration of the quantum geometry. We demonstrated then that this goes hand in hand with the accelerated and exponential expansion of its volume, as in Ref. \cite{GFCEmergentFriedmann2,GFCEmergentFriedmann3}.  The dynamics of the latter can be cast into the form of the classical Friedmann equations given in terms of the relational clock. In particular, we showed that the relative uncertainty of the volume operator vanishes
for such a configuration at late times which can be seen as an indication for the classicalization of the quantum geometry.

Solutions which avoid the singularity problem and grow exponentially after the bounce can only be found for negative ``GFT energy" in case of real-valued condensate fields. This is to be contrasted to the case of complex-valued fields where the mechanism responsible for the resolution of the initial singularity rests on the global $\mathrm{U}(1)$-symmetry of the field \cite{GFCEmergentFriedmann,GFCEmergentFriedmann2,GFCEmergentFriedmann3}.

We then moved on to study the formal stability properties of the evolving isotropic system when effective interactions are included. This was done using tools from the stability theory of general dynamical systems. In particular, we investigated the properties of solutions in the Thomas-Fermi regime. There, the system is dominated by the lowest-spin configuration of the quantum geometry and the dynamics of the volume can be cast into the form of modified Friedmann equations. Depending on whether the effective interaction potentials are bounded from below or not, the evolution of the emergent $3$-geometries will lead to a recollapse or to an infinite expansion, respectively. When studying in this regime the effect of two fine-tuned interaction terms mimicking those found in the GFT literature, by applying the techniques developed in Ref. \cite{GFCEmergentFriedmann3}, it is also possible to accommodate for an era of accelerated expansion which is strong enough to represent an alternative to the standard inflationary scenario. The study of the influence of effective interactions onto isotropic configurations was then concluded by performing a formal stability analysis of the full system including the effect of the Laplace-Beltrami operator. This allowed us to put the above given results into a larger context by showing that for a given value of the GFT ``mass", at small volumes low-spin modes are highly unstable, thus leading to their exponential growth over time. These are the physically most important modes, as they lead to a quick expansion of the emergent space.

Together with the results presented in Ref. \cite{GFClowspinStaticEffInt}, these results show that the condensate models considered here can give rise to effectively continuous, homogeneous and isotropic $3$-geometries built from many smallest and almost flat building blocks of the quantum geometry. Their dynamics lead to a rich phenomenology. In particular, it can be shown that the classical Friedmann equations may be recovered in an appropriate limit.

In a final step, we lifted the restriction of isotropy to study more general GFT condensate systems, thus opening an avenue to study anisotropies in the context of GFT condensate cosmology for the first time. Employing again formal stability analysis techniques known in the context of general dynamical systems, we showed for effectively interacting anisotropic configurations that the isotropic contribution with $j_1=j_2=j_3=1/2$ is highly unstable at small values of the relational clock, i.e. at small volumes. This quickly leads to the isotropization of the system for increasing values of the relational clock while the emergent geometry expands and thus suggests that within the GFT condensate cosmology framework a natural mechanism for smoothing out anisotropies may be realized. On the other hand, toward small volumes, we show that anisotropies surge. However, nothing like the anisotropy problem known in the literature on bouncing cosmologies occurs, where a regular bounce may be transformed into a singularity due to anisotropies, as reviewed in  Ref. \cite{Brandenberger}. We note that in our context the feature of singularity avoidance of solutions is not altered by the occurrence of anisotropies.\newline

\noindent In the following, we want to comment on the limitations and possible extensions of our discussion. 

Firstly, the work presented here is based on using real-valued GFT fields which to some extent leads to a different phenomenology compared to complex-valued fields since e.g. the singularity problem is avoided in different ways. To clarify this difference, a better understanding of the term ``GFT energy" would be desirable. Related to this is the question of how models exhibiting a bounce can be reconciled with the phase transition picture as a possible realization of the geometrogenesis scenario \cite{Geometrogenesis,GFTGeometrogenesis}. In the latter, the mean field is supposed to vanish at the critical point which cannot happen in models which exhibit bouncing solutions. In addition, it would be important to underpin the condensate conjecture by investigating whether a phase transition toward a condensate phase can be realized for GFT models on compact group manifolds or whether this is perhaps only realizable for models on noncompact domains like $\textrm{SL}(2,\mathbb{C})$, needed for Lorentzian quantum gravity models, as noticed in Ref. \cite{TGFTFRG, FRGcurvedspace}.
 
With regard to anisotropies, it would be interesting to repeat our analysis for complex-valued fields and study e.g. their impact on the bouncing mechanism found in Refs. \cite{GFCEmergentFriedmann,GFCEmergentFriedmann2}. Additionally, one should more systematically explore whether and how the dynamics of the volume can be cast into the form of generalized Friedmann equations for anisotropic cosmologies \cite{CosmologicalModels}. 

Given that the noncommutative Fourier transform can be easily computed from an individual solution, it would be important to reconstruct the corresponding metric from the quantum geometric information encapsulated by it and investigate its isometries. This would be to
some extent complementary to studying the dynamics of the volume computed for a particular model since the metric could tell us important information about both the geometry and topology of the $3$-space emerging from the condensate picture. In addition, the reconstruction of the metric would also be important in order to demonstrate that the coherent states used here indeed provide a suitable coarse-graining of the distributional quantum geometry such that it can be approximated by a smooth metric.

Furthermore, the impact of proper combinatorially nonlocal interactions on the model considered here should be investigated. In particular, studying the effect of simplicial interaction terms on the solutions and their formal stability would be highly interesting, since the quantum geometric interpretation is rather straightforward only for them. Related to this would be the extension of this approach beyond the Bogoliubov ansatz to have a better understanding of strongly interacting condensate systems.

Concerning the acceleration behavior of interacting GFT models, numerical techniques should be developed to generalize the results for more general configurations including anisotropies as well as for models going beyond the Thomas-Fermi regime as used here or in Refs. \cite{GFCEmergentFriedmann,GFCEmergentFriedmann2,GFCEmergentFriedmann3}. In particular, the robustness of the geometric inflation picture should be checked when going beyond the use of effective interactions.

Finally, in our analysis we made use of the notion of near-flatness in order to select mean field solutions corresponding to condensates consisting of almost flat building blocks of the quantum geometry. However, statements regarding the overall curvature of the emergent space can only be satisfactorily made if the expectation value of a currently unavailable GFT curvature operator is studied. This would most likely go along with exploring GFT condensates with more complicated building blocks such as dipoles. Studying these would also allow for a more direct comparison of the GFT condensate cosmology framework to the spin foam cosmology approach \cite{SFC} where those building blocks play a paradigmatic role. Such possible extensions are left to future research.

{\bf Acknowledgements.}
The authors thank A.~Ashtekar for remarks on classicalization criteria for the emergent geometry, S.~Gielen for several comments on an earlier version of this article, as well as D.~Oriti, M.~Schwoerer, L.~Sindoni and J.~Th\"urigen for useful discussions. The research of M.S. was supported in part by Perimeter Institute for Theoretical Physics. Research at Perimeter Institute is supported by the Government of Canada through the Department of Innovation, Science and Economic Development and by the Province of Ontario through
the Ministry of Research and Innovation.\newline

\begin{appendix}

\section{Noncommutative Fourier transform and reconstruction of the metric}\label{AppendixA}

In the following appendix, we review in what manner GFT states encode quantum geometric information dressing $3d$ simplicial complexes, in turn corresponding to quantum triangulations of spatial slices. We refer to \cite{GFC,GFCReview} for a more detailed account.

In the Hamiltonian formulation of Ashtekar-Barbero gravity, where $G=\textrm{SU}(2)$, the densitized inverse triad is canonically conjugate to the connection \cite{LQG}. The former represent momentum space variables in which the spatial metric can be written. Through a noncommutative Fourier transform which shifts between configuration and momentum space \cite{NCFT}, the GFT formalism can be dually formulated on the latter.

To this aim, consider the configuration space of the GFT field, i.e., $G^d$ with $d=4$ from which the phase space is constructed by the cotangent bundle $T^{*}G^4\cong G^4\times\mathfrak{g}^4$. The noncommutative Fourier transform of a square integrable GFT field is then defined by
\be\label{ncFT}
\tilde{\varphi}(B_I)=\int (dg)^4\prod_{I=1}^4 e_{g_{I}}(B_I)\varphi(g_I),
\ee
where the $B_I$ with $I=1,...,4$ denote the flux variables which parametrize the noncommutative momentum space $\mathfrak{g}^4$. The $e_{g_{I}}(B_I)$ represent a choice of plane waves on $G^4$. Their product is noncommutative, i.e., $e_{g}(B)\star e_{g'}(B)=e_{gg'}(B)$, indicated by the star product. Notice that we keep the vector arrows above the $B$s suppressed here. The noncommutative Dirac delta distribution on the momentum space is given by
\be
\delta_{\star}(B)=\int dg~ e_{g}(B).
\ee
With this object it is possible to show that the right invariance of the GFT fields corresponds to a closure condition for the fluxes, i.e., $\sum_I B_I=0$. It implies the closure of $I$ faces dual to the links $e_I$ to constitute a tetrahedron. Furthermore, it allows for the elimination of one of the $B_I$s when reexpressing the fluxes in terms of discrete triads. The latter is given by $B_i^{ab}=\int_{\triangle_i}e^a \wedge e^b$ with the cotriad field $e^a\in\mathbb{R}^3$ encoding the simplicial geometry and $\triangle_i$ with $i=1,2,3$ correspond to the faces associated to the tetrahedron. As shown in Ref. \cite{GFC}, the metric at a given fixed point in the tetrahedron can be reconstructed from this by means of
\be\label{reconstructedmetric}
g_{ij}=e^a_i e^b_j\delta_{ab}=\frac{1}{4 \tr(B_1 B_2 B_3)}\epsilon_i^{kl}\epsilon_{j}^{mn}\tilde{B}_{km}\tilde{B}_{ln},
\ee
where $\tilde{B}_{ij}\equiv \tr(B_iB_j)$ holds. 

When promoting the fields in Eq. (\ref{ncFT}) to field operators, the metric of a quantum tetrahedron is determined by $\hat{\tilde{{\varphi}}}^{\dagger}(B_i)|\emptyset\rangle=|B_i\rangle$. This also holds true for the momentum space representation of the condensate mean field $\sigma(g_I)$.

\end{appendix}

\end{document}